\documentclass[preprint,nofootinbib]{revtex4}
\usepackage{grffile}
\usepackage{graphicx}
\newcommand{\pt}{p_{\rm T}} 
\newcommand{\snn}{\sqrt {s_{\rm NN}}}
\newcommand{\np}{N_{\rm part}}

\begin{document}
\title{Predictions for  $\sqrt {s_{NN}}=5.02$ TeV Pb+Pb
  Collisions from a Multi-Phase Transport Model}
\author{Guo-Liang Ma}
\affiliation{Shanghai Institute of Applied Physics, Chinese Academy of Sciences, Shanghai 201800, China}
\email[]{glma@sinap.ac.cn}
\author{Zi-Wei Lin}
\affiliation{Department of Physics, East Carolina University,
  Greenville, North Carolina 27858, USA}
\email[]{linz@ecu.edu}

\begin{abstract}
We present predictions from the string melting version of a
multi-phase transport model on various observables in Pb+Pb
collisions at $\sqrt {s_{NN}}=5.02$ TeV. 
We use the same version of the model as an earlier study that
reasonably reproduced dN/dy, $p_{\rm T}$-spectra and 
elliptic flow of charged pions and kaons at low-$p_{\rm T}$ for central and
semi-central heavy ion collisions at 200 GeV and 2.76 TeV. 
While we compare with the already-available centrality dependence data
on charged particle $dN/d\eta$ at mid-pseudorapidity
in Pb+Pb collisions at 5.02 TeV, we make predictions on identified
particle dN/dy, $p_{\rm T}$-spectra, azimuthal anisotropies $v_n (n=2,3,4)$,
and factorization ratios $r_{n}(\eta^{a},\eta^{b}) (n=2,3)$ for
longitudinal correlations. 
\end{abstract}
\maketitle

\section{Introduction}

Ultra-relativistic heavy ion collisions at the Relativistic Heavy Ion
Collider (RHIC) and the Large Hadron Collider (LHC) have created a
dense matter consisting of partonic degrees of freedom, often called
the quark-gluon plasma (QGP).
From comparisons between the experimental data on various observables
and results of different theoretical models, we expect to learn the
space-time evolution of the dense matter and consequently the
properties of QGP and quantum chromodynamics. 
For collisions of heavy ions such as Pb+Pb, the highest energy
achievable in the near future is $\snn=5.02$ TeV. 
Since high energy heavy ion collisions are expected to create a QGP
matter with a high initial effective temperature with a long
lifetime being in parton degrees of freedom, comparing observables at
this highest energy with those at lower energies will provide us new
information for the study of QGP properties. 

Simulations of the space-time evolution of relativistic heavy ion
collisions have been performed extensively with hydrodynamic models
\cite{Huovinen:2001cy,Betz:2008ka,Schenke:2010rr}, 
transport models \cite{Xu:2004mz,Lin:2004en,Cassing:2009vt}, and
hybrid models that combine a hydrodynamic model with a transport model
\cite{Petersen:2008dd,Werner:2010aa,Song:2010mg}. 
Despite the different physics foundations and assumptions in these
three types of models, they have been quite successful in describing
many observables such as azimuthal anisotropies. 
It has been widely believed that for heavy colliding systems
transport models essentially approach the hydrodynamical limit, 
therefore it does not seem surprising that transport and hydrodynamic models
can both describe anisotropy observables in large systems. 
On the other hand, it is puzzling that they also both seem to describe
azimuthal anisotropies in small systems such as p+Pb and d+Au collisions
\cite{Bozek:2011if,Ma:2014pva,Bzdak:2014dia}, since transport models are far
from the hydrodynamical limit for small systems. 

Recently it was realized that transport models for current
ultra-relativistic heavy colliding systems may still be far 
from the hydrodynamical limit. In particular, it is
found with transport models \cite{He:2015hfa,Lin:2015ucn} that
azimuthal anisotropies may be produced mainly by the anisotropic
parton escape probability as a response to the initial spatial
eccentricity through interactions, not mainly by the hydrodynamic-type
collective flow. Therefore transport models and 
hydrodynamic models for current heavy ion collisions are
different, and it is important to identify unique signatures of
transport or hydrodynamic models in order to determine which picture is a more
relevant description for a given type of heavy ion collisions. 
There have been recent hydrodynamic predictions for 5.02 TeV Pb+Pb
collisions \cite{Niemi:2015voa,Noronha-Hostler:2015uye}, and in this
study we present transport predictions from a multi-phase transport
(AMPT) model \cite{Lin:2004en}. 

\section{The string melting version of the AMPT model and its
  parameters} 

The AMPT model was constructed to 
simulate relativistic heavy ion collisions, and predictions were made
for Au+Au collisions at RHIC energies \cite{Zhang:1999bd, Bass:1999zq,
  Lin:2000cx}. 
The model at that time (now called the default version) consisted of  
fluctuating initial conditions from the HIJING model \cite{Gyulassy:1994ew}, 
the elastic parton cascade ZPC \cite{Zhang:1997ej} for minijet partons,  
the Lund string model \cite{Sjostrand:1993yb} for hadronization,  
and the ART hadron cascade \cite{Li:1995pra}. 
With modified Lund string fragmentation $a$ and $b$ parameters
\cite{Lin:2000cx,Lin:2004en}, the default AMPT model was able to 
reasonably describe the rapidity distributions and $\pt$ spectra in
heavy ion collisions from CERN Super Proton Synchrotron (SPS) to RHIC energies. 

It was later found that the default version of the AMPT model (as all transport
models at the time) under-estimated the elliptic flow observed at
RHIC, and the reason was that most of the energy produced in the
overlap volume of heavy ion collisions are in hadronic strings and
thus not included in the parton cascade in the model \cite{Lin:2001zk}. 
So the string melting version of the AMPT model 
was constructed, where all excited hadronic strings in the overlap volume
are converted into partons \cite{Lin:2001zk}.
The string melting AMPT model consists of fluctuating initial conditions from the
HIJING model \cite{Gyulassy:1994ew}, the elastic parton cascade ZPC
\cite{Zhang:1997ej} for all partons 
from the melting of hadronic strings, a quark coalescence model for
hadronization, and the ART hadron cascade \cite{Li:1995pra}.
Due to its dense parton phase,
the string melting version reasonably fit the elliptic flow 
\cite{Lin:2001zk} and two-pion interferometry \cite{Lin:2002gc}
in heavy ion collisions at RHIC energies; but it could not reproduce
well the rapidity distributions and $\pt$ spectra (when using the same
Lund $a$ and $b$ parameters as in the default version). 
Therefore, in previous predictions from the AMPT model on Pb+Pb
collisions at 5.5 TeV \cite{Abreu:2007kv}, we used  the default AMPT
model to predict particle yields and $\pt$ spectra but used  the
string melting AMPT model to predict the elliptic flow and
two-pion or two-kaon correlation functions. 

Recently we found that the string melting AMPT model can be tuned
to reasonably reproduce the pion and kaon yields, $\pt$ spectra, 
and elliptic flows below $\sim 1.5$ GeV/$c$ in central
and semi-central Au+Au collisions at the RHIC energy of 
$\snn=200$ GeV and Pb+Pb collisions at the LHC energy of 2.76 TeV
\cite{Lin:2014tya}.  
For predictions for the top LHC energy of 5.02 TeV in this study,
we use the same string melting AMPT version (v2.26t5, available online
\cite{ampt}) and the same parameter values as used for the LHC energy
of 2.76 TeV in an earlier study \cite{Lin:2014tya}. 
These parameters include the Lund string fragmentation parameters
$a=0.30$ and $b=0.15$ GeV$^{-2}$, strong coupling constant
$\alpha_s=0.33$, and a parton cross section of 3 mb. Note that this
AMPT version \cite{Lin:2014tya,ampt} imposes an upper limit of 0.40
on the relative production of strange to nonstrange quarks from the
Lund string fragmentation.  
In addition, in order to avoid potential effects due to different Lund
$a$-values, we also use these parameters (Lund $a=0.30$ in particular) 
for the 200 GeV RHIC simulations in this study, even though they
describe the 200 GeV dN/dy data not as well as the earlier study
\cite{Lin:2014tya} which used Lund $a=0.55$. 
Since our predictions include observables far away from
mid-(pseudo)rapidity, we terminate the hadron cascade at 200 fm/$c$ 
in all AMPT calculations of this study instead of the typical value of
30 fm/$c$. Note that centrality in AMPT results of this study is
determined by the range of impact parameters in the simulated
minimum-bias events  
for which we impose no upper limit on the maximum impact parameter. 
For example, for Pb+Pb collisions at 5.02 TeV
the 0-5\% centrality  corresponds to impact parameters from 0 to 3.6
fm while the 20-30\% centrality corresponds to impact parameters from
7.3 fm to 8.9 fm.

\section{Comparison of AMPT results with the available $dN_{ch}/d\eta$ data}

\begin{figure}[h]
\includegraphics[width=6 in]{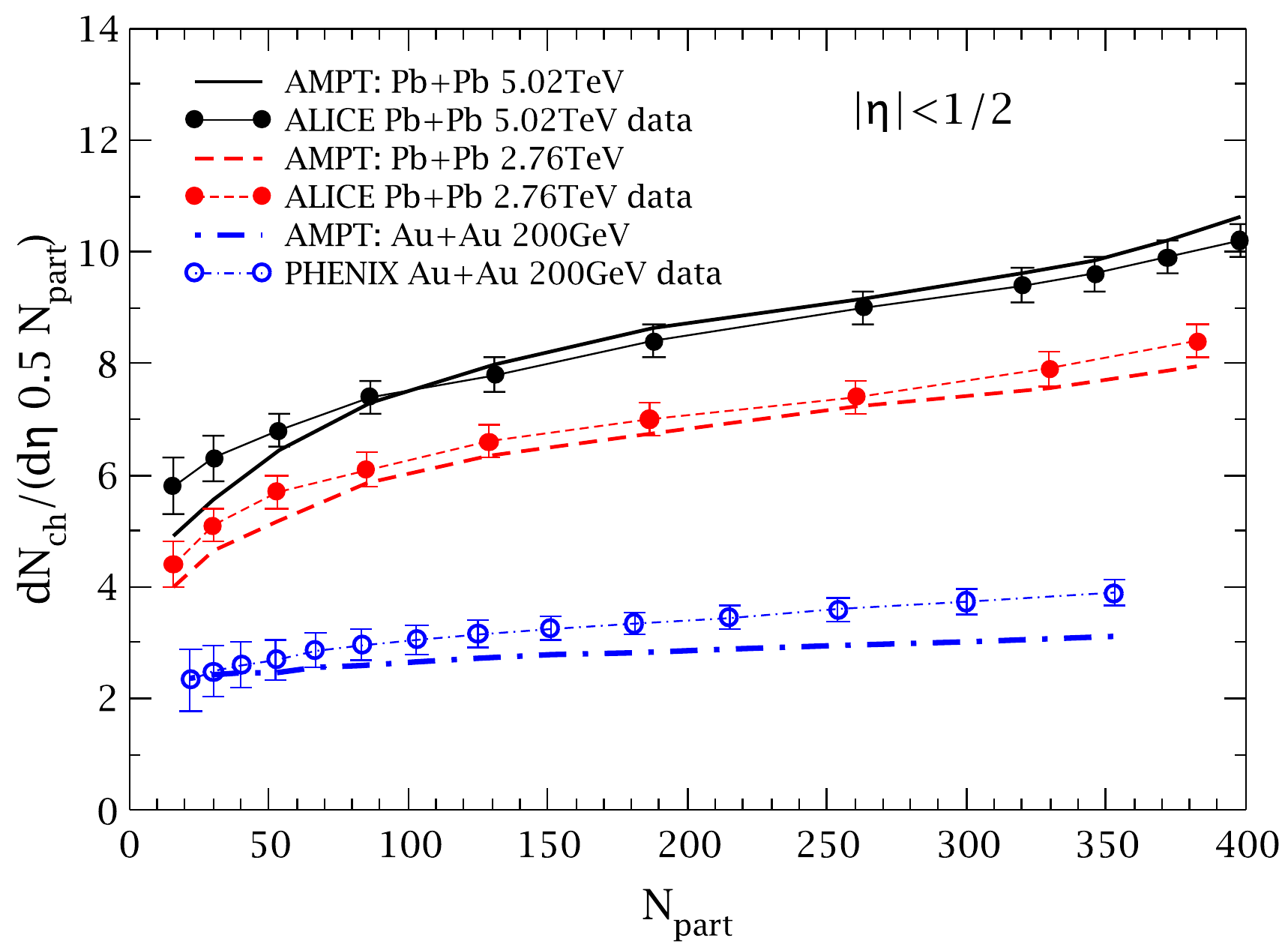}
\caption{The centrality dependence
  of $dN_{ch}/d\eta$ at mid-pseudorapidity ($|\eta|<1/2$) 
from the string melting AMPT model in comparison with data 
for heavy ion collisions at $\snn=5.02$ TeV, 2.76 TeV and 200 GeV.} 
\label{fig:dnchde}
\end{figure}

We compare in Fig.\ref{fig:dnchde} charged particle yields at
mid-pseudorapidity, scaled by half the number of participant nucleons $\np$,
from the string melting AMPT model (curves without
symbols) in comparison with the experimental data, which 
include the ALICE Pb+Pb data at $\snn=5.02$ TeV \cite{Adam:2015ptt}, 
2.76 TeV \cite{Aamodt:2010cz}, and the PHENIX Au+Au data at 200 GeV
\cite{Adler:2004zn}.
Although these AMPT results were already obtained with
the same AMPT code and parameters as in an earlier study
\cite{Lin:2014tya} before the announcement of the ALICE 5.02 TeV data
\cite{Adam:2015ptt}, they were not posted before the data and are
thus, strictly speaking, not predictions. 

\begin{figure}[h]
\includegraphics[width=6 in]{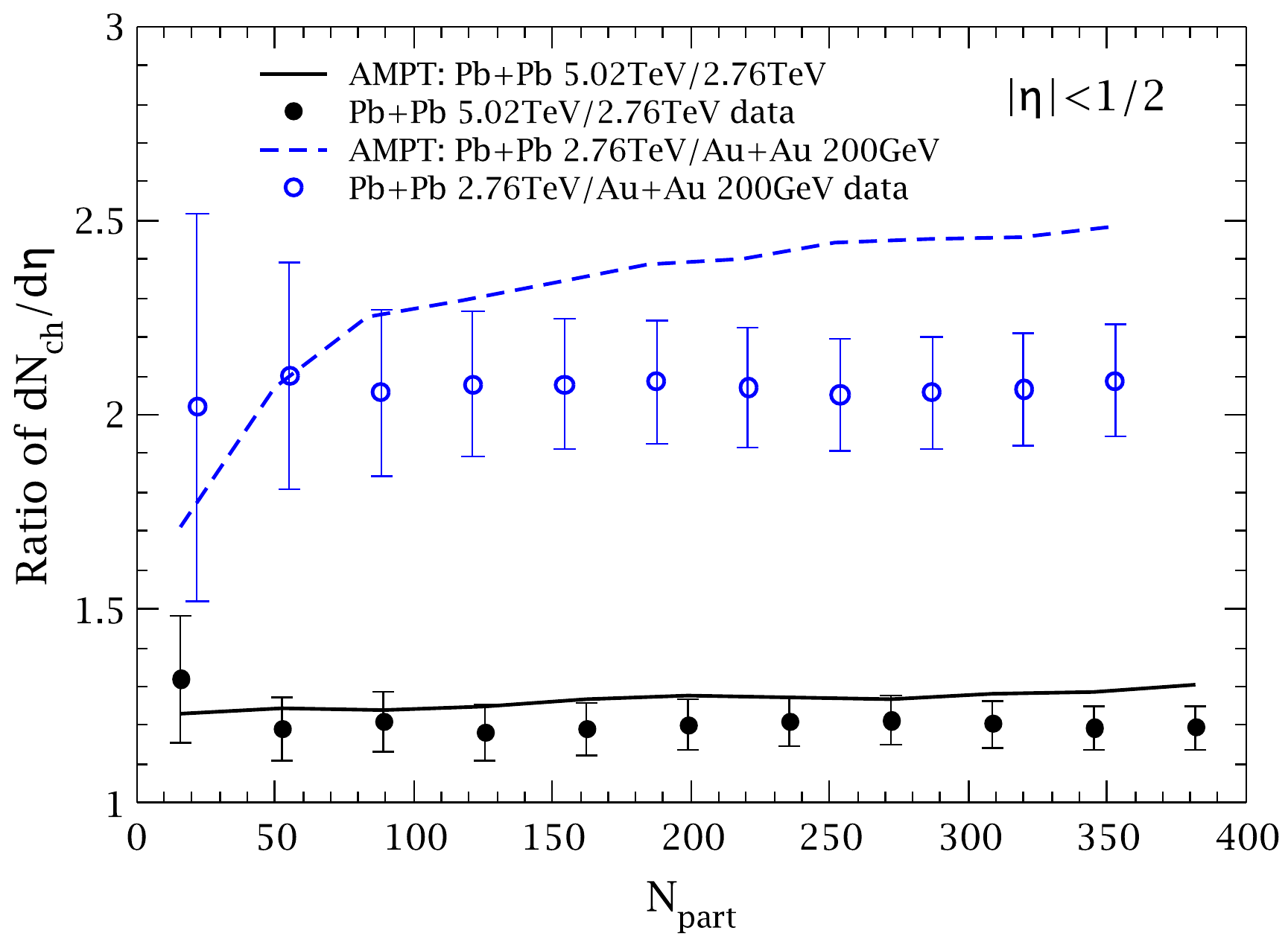}
\caption{Ratio of mid-pseudorapidity $dN_{ch}/d\eta$ at
  two different energies from AMPT in comparison with 
the corresponding ratio derived from experimental data.}
\label{fig:dnchderatio}
\end{figure}

Figure~\ref{fig:dnchde} shows that the AMPT model reproduces the overall
magnitudes and the $\np$ dependence shapes of $dN_{ch}/d\eta$ at 
mid-pseudorapidity ($|\eta|<1/2$). 
Note that the AMPT 200 GeV results in this study are obtained using
the same Lund $a$ parameter ($a=0.30$) as the LHC results; 
they are generally lower than the PHENIX data
and also lower than the AMPT 200 GeV results 
in an earlier study \cite{Lin:2014tya} where the Lund parameter
$a=0.55$ was used. 
We also see in Fig.\ref{fig:dnchde} that the energy dependence of
$dN_{ch}/d\eta$ at mid-pseudorapidity from
AMPT is often stronger than the data. 
This can be better demonstrated in Fig.\ref{fig:dnchderatio}, 
where we show the ratio of mid-pseudorapidity $dN_{ch}/d\eta$
at one energy to that at a lower energy. 
The ratio of mid-pseudorapidity $dN_{ch}/d\eta$ in Pb+Pb collisions at
2.76 TeV to  that in Au+Au collisions at 200 GeV from AMPT (dashed
curve) is usually higher than the corresponding data (open
circles), where the data represent the ratio of the ALICE data for Pb+Pb
collisions at 2.76 TeV to the PHENIX data for Au+Au collisions at
200 GeV. 
On the other hand, the ratio of mid-pseudorapidity $dN_{ch}/d\eta$ in
Pb+Pb collisions at 5.02 TeV to that at 2.76 TeV from AMPT (solid
curve) is close to the corresponding data,  
and its magnitude is much lower than the other ratio (dashed curve) 
as a result of the smaller relative increase of the collision energy. 
Note that, since the experimental data (and AMPT results) in
Fig.\ref{fig:dnchde} at different energies have different sets of $\np$
values, we have used linear interpolation in order to make the
$dN_{ch}/d\eta$ ratio at a given $\np$ value for
Fig.\ref{fig:dnchderatio}.

\begin{figure}[h]
\includegraphics[width=6 in]{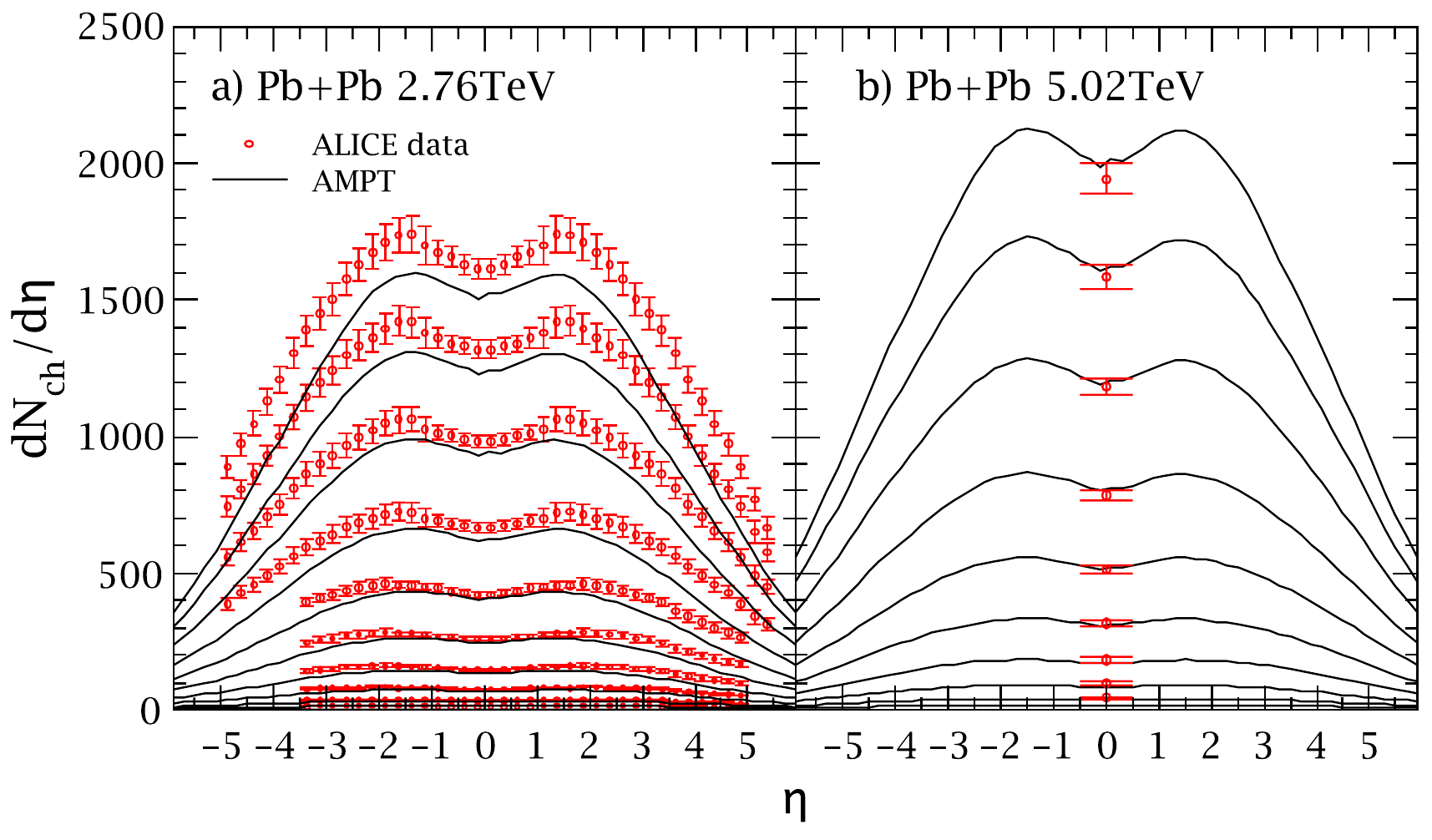}
\caption{
$dN_{ch}/d\eta$ from the string melting AMPT model 
for Pb+Pb collisions at a) $\snn=2.76$ TeV and
b) 5.02 TeV in comparison with data (circles). Curves from top to
bottom represent respectively the AMPT results for the following
centralities: 0-5\%, 5-10\%, 10-20\%, 20-30\%, 30-40\%, 40-50\%,
50-60\%, 60-70\%, 70-80\%, and 80-90\%.
} 
\label{fig:dnchdeta}
\end{figure}

The $dN_{ch}/d\eta$ results as
functions of $\eta$ at different centralities of Pb+Pb collisions 
are shown in Fig.~\ref{fig:dnchdeta}a) for 
$\snn=2.76$ TeV and in Fig.~\ref{fig:dnchdeta}b) for 5.02 TeV.
Solid curves from top to bottom in each panel represent respectively
the AMPT results for the following centralities: 0-5\%, 5-10\%,
10-20\%, 20-30\%, 30-40\%, 40-50\%, 50-60\%, 60-70\%, 70-80\%, and
80-90\%;  while circles represent the corresponding ALICE data 
\cite{Abbas:2013bpa,Adam:2015kda,Adam:2015ptt}.   
Note that the ALICE data at 5.02 TeV \cite{Adam:2015ptt} represent the
$dN_{ch}/d\eta$ values at mid-pseudorapidity ($|\eta|<1/2$) and they
do not include the 80-90\% centrality.
We see that the AMPT model reproduces the overall 
$dN_{ch}/d\eta$ shape, but the AMPT $dN_{ch}/d\eta$ curves at 2.76 TeV
are a bit narrower than the corresponding data.

\section{AMPT Predictions} 

This section shows our predictions on dN/dy and $\pt$-spectra of
identified particles including $\pi^+$ and $K^+$, 
azimuthal anisotropies including the $\pt$ dependences of $v_n
(n=2,3,4)$ and $\eta$ dependences of $v_2$, and the factorization
ratios $r_{2}(\eta^{a},\eta^{b})$ and $r_{3}(\eta^{a},\eta^{b})$ for
longitudinal correlations.  

\subsection{Identified Particle dN/dy}

\begin{figure}[h]
\includegraphics[width=6 in]{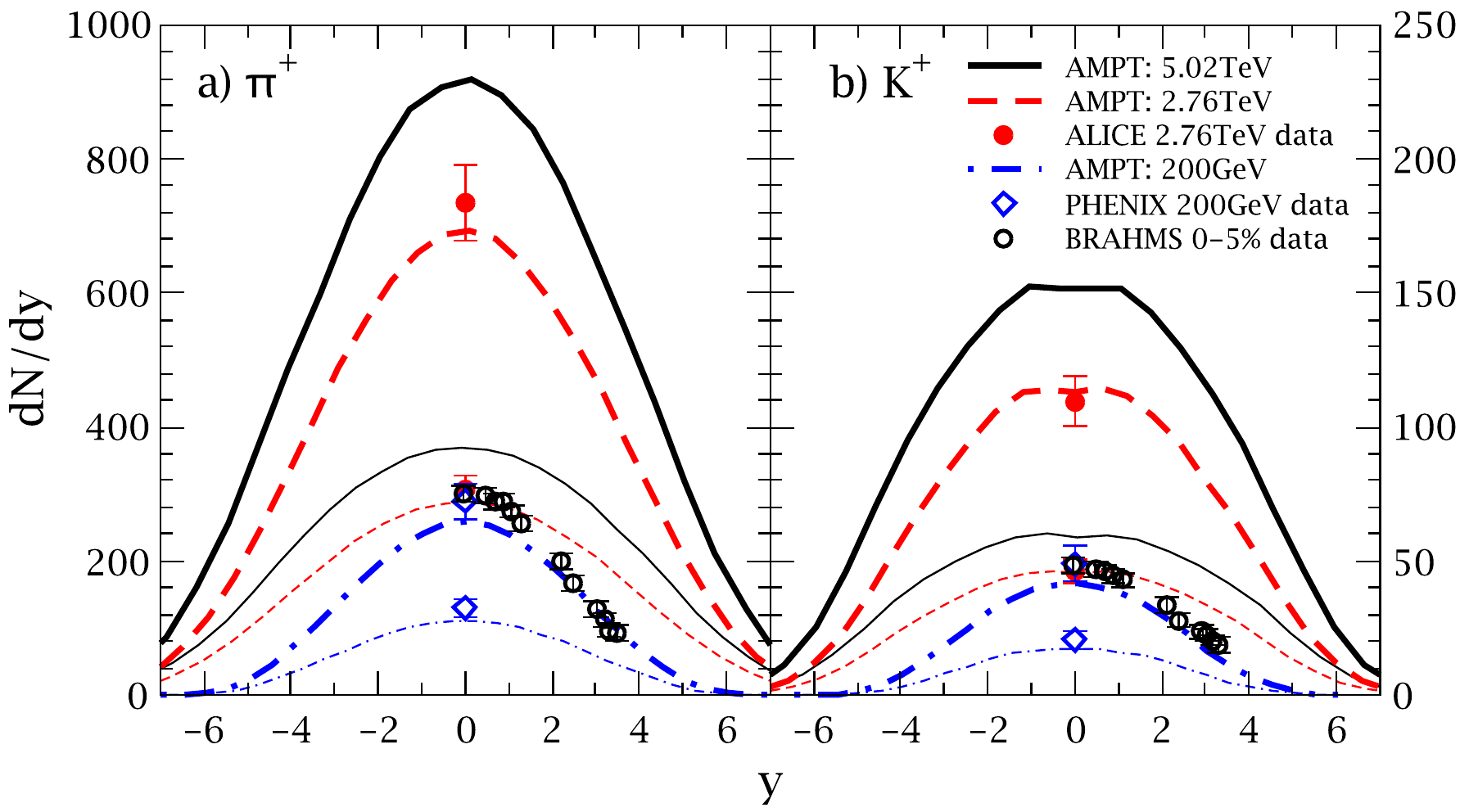}
\caption{ 
AMPT results on dN/dy of a) $\pi^+$ and b) $K^+$
for Pb+Pb collisions at 5.02 TeV, 2.76 TeV and 
Au+Au collisions at 200 GeV, in comparison with
experimental data at 2.76 TeV and 200 GeV, for 0-5\% and 20-30\%
centralities. 
}
\label{fig:dndy}
\end{figure}

AMPT results of $\pi^+$ and $K^+$ dN/dy are shown 
in Figs.\ref{fig:dndy}a and \ref{fig:dndy}b, respectively, 
where thick curves represent the 0-5\% centrality and 
thin curves represent the 20-30\% centrality. 
Experimental data for Pb+Pb collisions at 2.76 TeV (filled symbols) 
and Au+Au collisions at 200 GeV (open symbols) are also shown for
comparison with the corresponding AMPT results (dashed curves for 2.76
TeV and dot-dashed curves for 200 GeV).  
We see reasonable agreements between the AMPT results and
the mid-rapidity ALICE data at 2.76 TeV \cite{Abelev:2013vea} and 
PHENIX data at 200 GeV \cite{Adler:2003cb}  for both centralities.
Reasonable agreements are also seen in comparison with the 
BRAHMS rapidity dependence data for 0-5\% central Au+Au collisions at
200 GeV \cite{Bearden:2004yx}. Note that the pion 
and kaon yields from PHENIX and ALICE have been corrected for weak
decays (especially those of strange baryons) and can thus be directly
compared with AMPT results. 

When the energy of Pb+Pb collisions increases from 2.76 TeV to 5.02
TeV,  we see from Fig.\ref{fig:dndy} that the $\pi^+$ dN/dy at
mid-rapidity increases by $\sim 33\%$ for the 0-5\% centrality   
and $\sim 28\%$ for the 20-30\% centrality,  
while the increase for $K^+$ is 
$\sim 34\%$ for the 0-5\% centrality 
and $\sim 28\%$ for the 20-30\% centrality. 

\subsection{Identified Particle $\pt$-Spectra}

\begin{figure}[h]
\includegraphics[width=6 in]{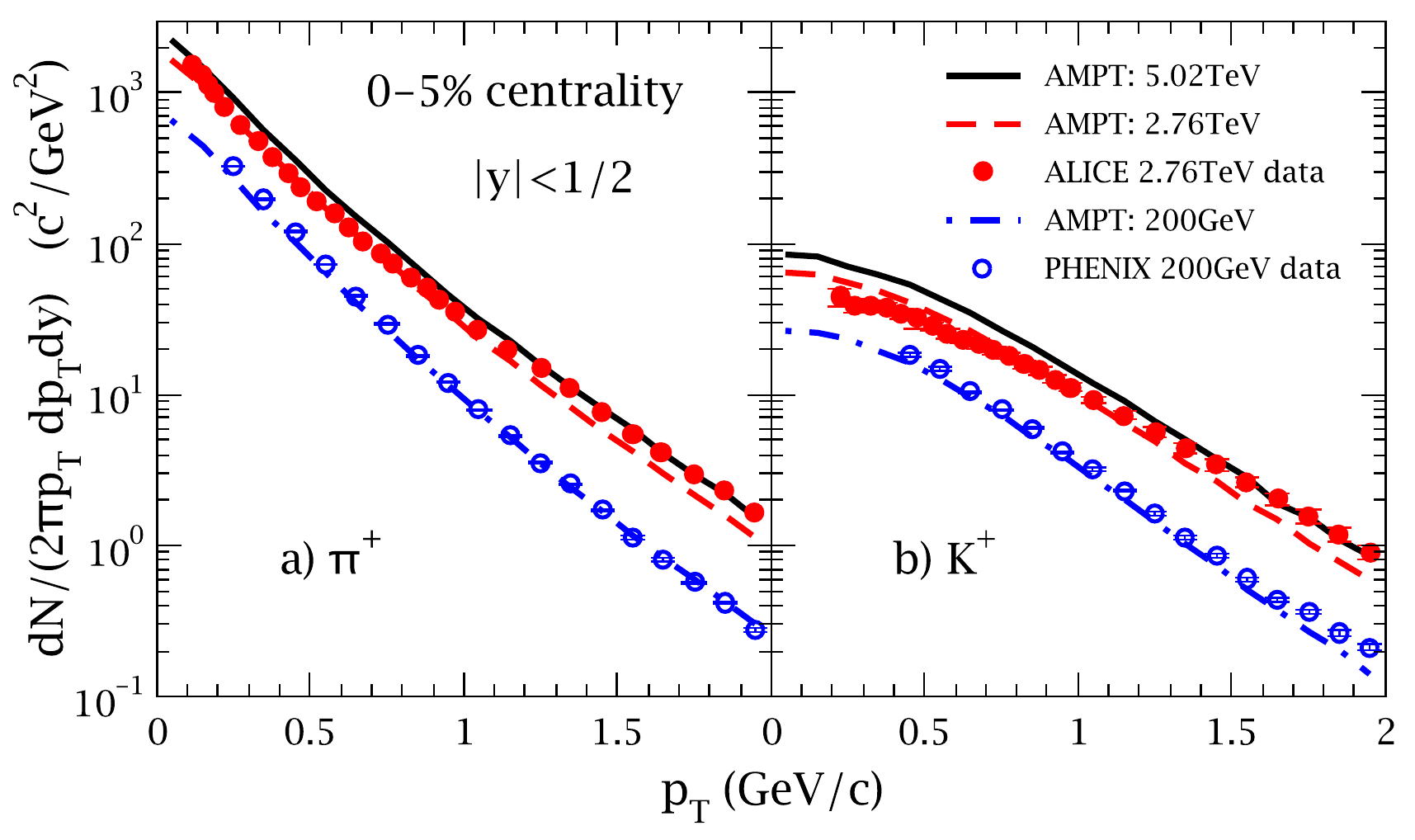}
\caption{ 
AMPT results on a) $\pi^+$ and b) $K^+$ $\pt$ spectra at mid-rapidity 
in Pb+Pb collisions at 5.02 TeV, 2.76 TeV, and 
Au+Au collisions at 200 GeV, in comparison with available
experimental data, for the 0-5\% centrality.
}
\label{fig:pt}
\end{figure}

\begin{figure}[h]
\includegraphics[width=6 in]{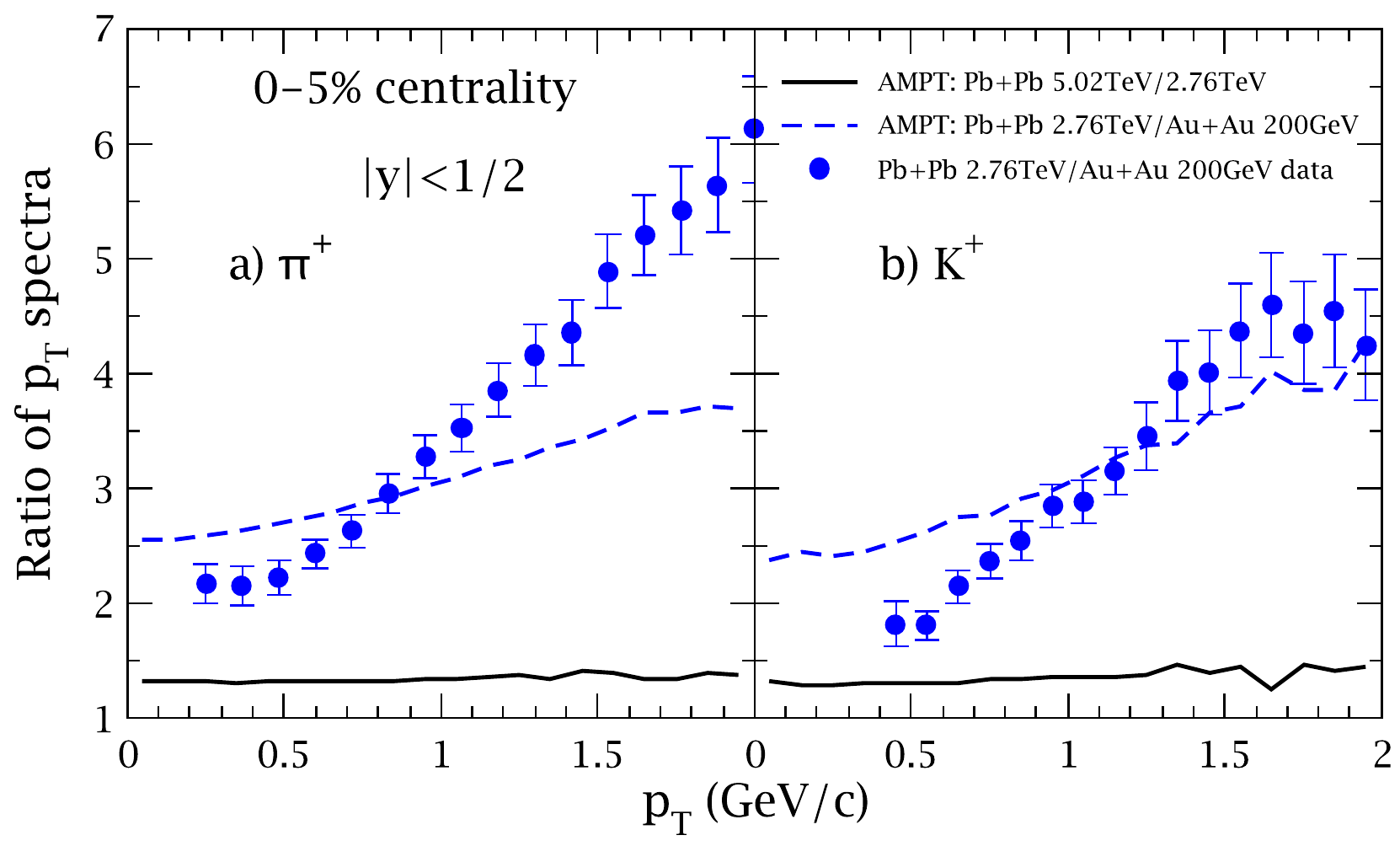}
\caption{Ratio of $\pt$ spectra at
  mid-rapidity at two different energies from AMPT in
  comparison with data for the 0-5\% centrality 
for a) $\pi^+$ and b) $K^+$.}
\label{fig:ptratio}
\end{figure}

AMPT results on the $\pi^+$ and $K^+$ $\pt$ spectra are shown in
Fig.\ref{fig:pt} for 0-5\% central Pb+Pb collisions at 5.02 TeV (solid
curves), 2.76 TeV (dashed curves), and 0-5\% Au+Au collisions at
200 GeV (dot-dashed curves). 
The 0-5\% ALICE data at 2.76 TeV \cite{Abelev:2013vea} and 0-5\%
PHENIX data at 200 GeV \cite{Adler:2003cb} are shown by symbols. 
We see that the AMPT model roughly reproduces the 
observed $\pt$ spectra at low $\pt$ at the lower two energies. 

To better observe the change of the $\pt$ spectra, we show in 
Fig.\ref{fig:ptratio} the ratio of the mid-rapidity $\pt$ spectrum 
at one energy to that at a lower energy. 
The AMPT ratio of Pb+Pb collisions at 5.02 TeV to those at 2.76 TeV
(solid curves) shows a weak increase with $\pt$ (from $\sim 1.3$ at
$\pt=0$ to  $\sim 1.4$ at $\pt=2$ GeV/$c$) for both $\pi^+$ and $K^+$, 
indicating that the $\pt$ spectrum becomes slightly harder with the
increase in energy. 
In comparison, the hardening of the $\pt$ spectra as well as the
overall increase in magnitude are much stronger going from central
Au+Au collisions at 200 GeV to central Pb+Pb collisions at 2.76 TeV. 
However, the AMPT results (dashed curves) 
under-estimate the hardening of the $\pt$ spectra when compared with   
the experimental data (circles). 

\subsection{Azimuthal Anisotropies}

Azimuthal anisotropies such as $v_2$,  $v_3$ and $v_4$ 
reflect the response to the initial spatial eccentricities, which are
generated by the overall geometry in non-central collisions and/or
fluctuations, through interactions
during the evolution of the dense matter. 
Detailed studies of these anisotropic flow observables enable us
to study whether the system is close to 
hydrodynamics \cite{He:2015hfa,Lin:2015ucn}
and to extract the QGP properties
\cite{Huovinen:2001cy,Betz:2008ka,Schenke:2010rr} if the system is
close to the hydrodynamic limit.

\begin{figure}[h]
\includegraphics[width=6 in]{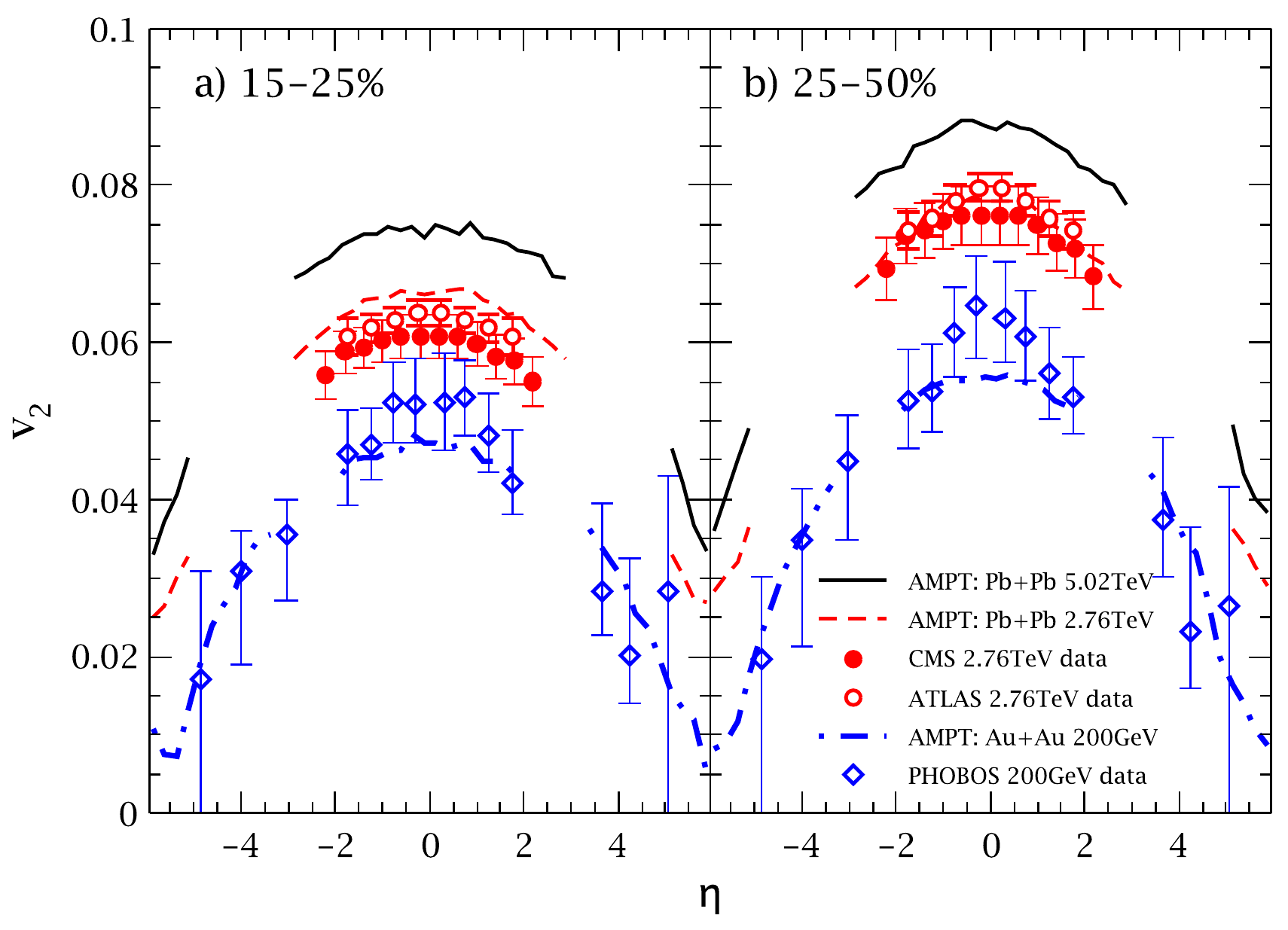}
\caption{AMPT results on the $\eta$ dependence of
  $v_2$, in comparison with available experimental data,  
for a) the 15-25\% centrality and b) the 25-50\% centrality.
}
\label{fig:v2eta}
\end{figure}

\begin{figure}[h]
\includegraphics[width=6 in]{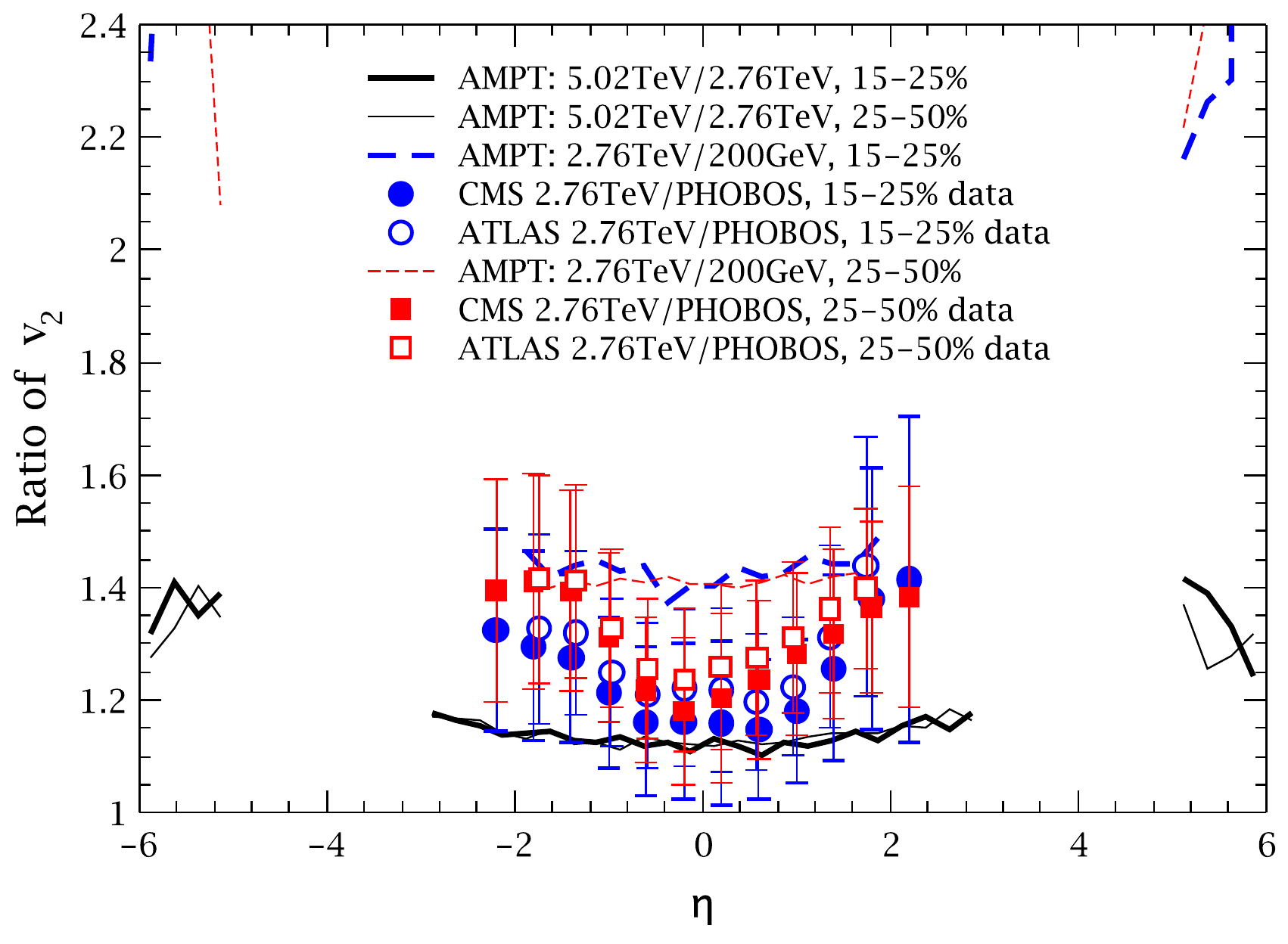}
\caption{Ratio of $v_2(\eta)$ at two different energies 
from AMPT in comparison with available experimental data.
}
\label{fig:v2etaratio}
\end{figure}

In this study we use two methods to calculate the azimuthal
anisotropies. For Au+Au collisions at 200 GeV, we use the Event-Plane
method to calculate $v_n$\{EP\}, in order to compare with the PHENIX
data. We calculate the $n$-th event plane $\Psi_{n}$ based on the
momentum information of final particles \cite{Poskanzer:1998yz}:
\begin{equation} \label{psi}
\Psi_{n}=\frac{1}{n}\left ({\rm arctan} 
\frac{\left\langle
      p_T \sin (n\phi)\right\rangle} {\left\langle
      p_T \cos (n\phi)\right\rangle}  \right), 
\end{equation}
where $\phi$ represents the azimuthal angle of a particle's momentum,
and $\langle \cdots\rangle$ denotes the average 
over particles used for the event plane calculation.
Then the $n$-th anisotropy coefficient, $v_n$\{EP\}, can be
obtained as 
\begin{equation} \label{vn}
v_n\{\rm EP\}=\left\langle \cos \left[ n(\phi-\Psi_{n}) \right] \right\rangle/ Res\{\Psi_{n}\},
\end{equation}
where Res\{$\Psi_{n}$\} is the $n$-th event plane resolution
calculated by the experimental subevent method. We use all particles
within the acceptance of PHENIX event plane detectors,
$1.0<|\eta|<2.8$,  to reconstruct the event plane. 
For Pb+Pb collisions at 2.76 and 5.02 TeV, we use 
the two-particle method with a large $\Delta\eta$ gap to calculate
$v_n$\{2\}, similar to the CMS \cite{Chatrchyan:2013nka} and the ATLAS
\cite{ATLAS:2012at} methods 
[except for $v_2(\eta)$ calculations as shown in
Figs.\ref{fig:v2eta}-\ref{fig:v2etaratio} where we use the Event-Plane
method].
The two-particle azimuthal $\Delta\phi$ correlation function is
decomposed as 
\begin{equation}
\label{Vn}
\frac{1}{N_{\rm trig}}\frac{d N^{\rm pair}}{d\Delta\phi} =
\frac{N_{\rm assoc}}{2\pi} \left[ 1+\sum\limits_{n} 2V_{n\Delta} \cos
  (n\Delta\phi)\right], 
\end{equation}
where $V_{n\Delta}$ is the Fourier coefficient and $N_{\rm assoc}$
represents the total number of pairs per trigger particle. A cut of
$|\Delta\eta|>$2 is applied to remove short-range correlations from
jet fragmentation for charged particles within $|\eta|<$2.5.  The
azimuthal anisotropy coefficients, $v_{n}\{2,|\Delta\eta|>2\}$, from the
two-particle correlation method can be extracted as a function of
$\pt$ from the fitted Fourier coefficients as 
\begin{equation}
\label{vnpt}
v_{n}\{2,|\Delta\eta|>2\}(\pt) = \frac{V_{n\Delta}(\pt,\pt^{\rm
    ref})}{\sqrt{V_{n\Delta}(\pt^{\rm ref},\pt^{\rm ref})}}, 
\end{equation}
where a fixed $\pt^{\rm ref}$ range for the ``reference particles'' is
chosen to be $0.3<\pt<3.0$ GeV/$c$ in our study.  It have been found
that the two methods give very similar results
\cite{ATLAS:2012at,Gu:2012br}.  

\begin{figure}[h]
\includegraphics[width=6 in]{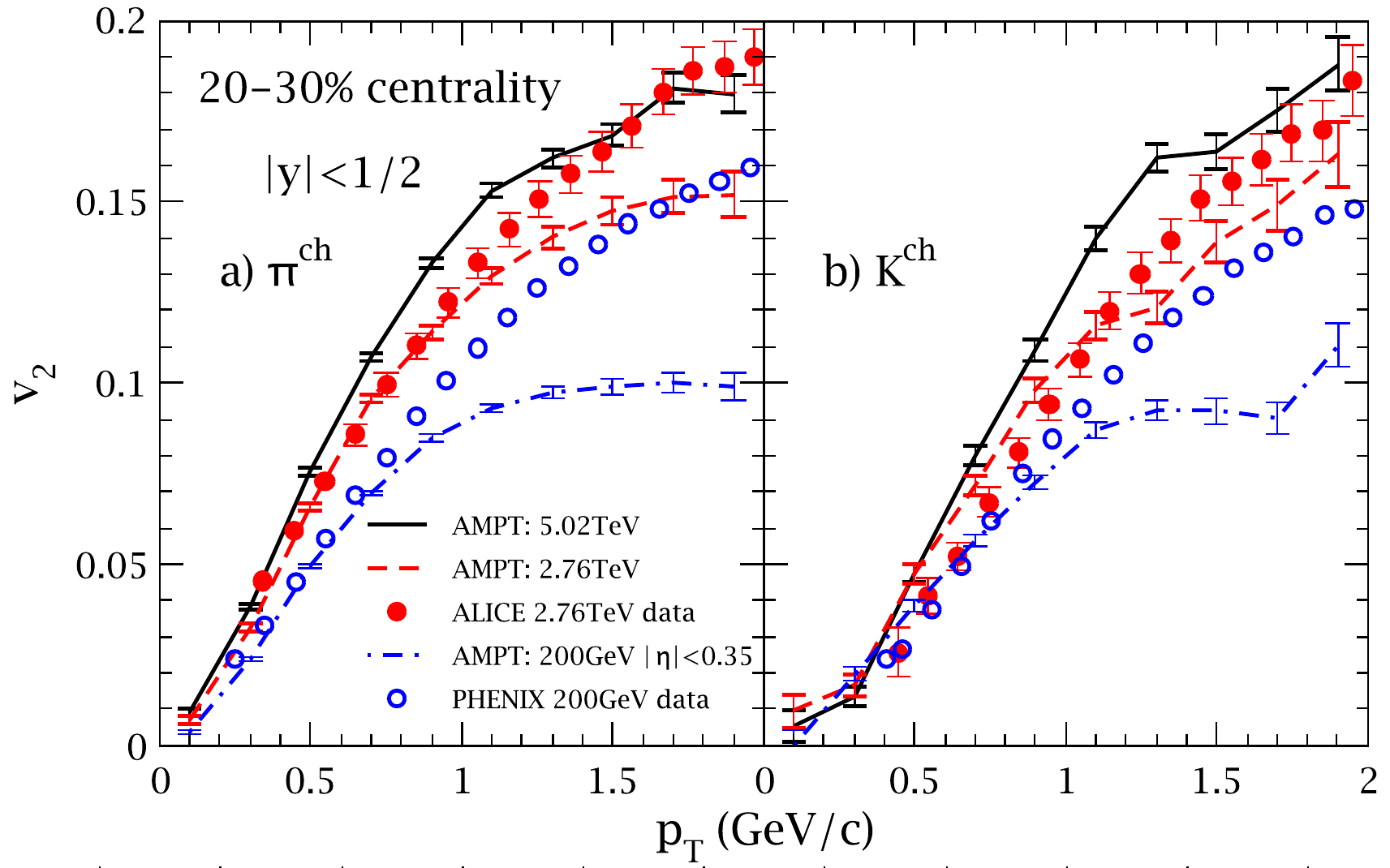}
\caption{ 
The $\pt$ dependence of $v_2$ at mid-rapidity from AMPT, 
in comparison with available experimental data, for the
20-30\% centrality for a) charged pions and b) charged kaons.
}
\label{fig:v2}
\end{figure}

AMPT results on the $\pt$ dependence of $v_2$ at mid-rapidity are
shown in Fig.\ref{fig:v2}a for charged pions and 
Fig.\ref{fig:v2}b for charged kaons, 
in comparison with the ALICE data for Pb+Pb collisions at 2.76 TeV 
(filled circles) and PHENIX data for Au+Au collisions at 200 GeV
(open circles) \cite{Gu:2012br}, for the 20-30\% centrality.
The 200 GeV AMPT results has the cut $|\eta|<0.35$
for comparison with the PHENIX data. 
We see that the AMPT model reasonably reproduces the $v_2$ data 
below $\pt \sim 1$ GeV/$c$ at these two energies.
For Pb+Pb collisions at 5.02 TeV in comparison with those at 2.76 TeV,
the AMPT model predicts that the pion $v_2$ has a relative increase of
$\sim 16\%$ that is rather insensitive to $\pt$ within 
0 and 2 GeV/$c$, while the relative increase of the kaon $v_2$
shows an overall increase with $\pt$ within this $\pt$ range.

\begin{figure}[h]
\includegraphics[width=6 in]{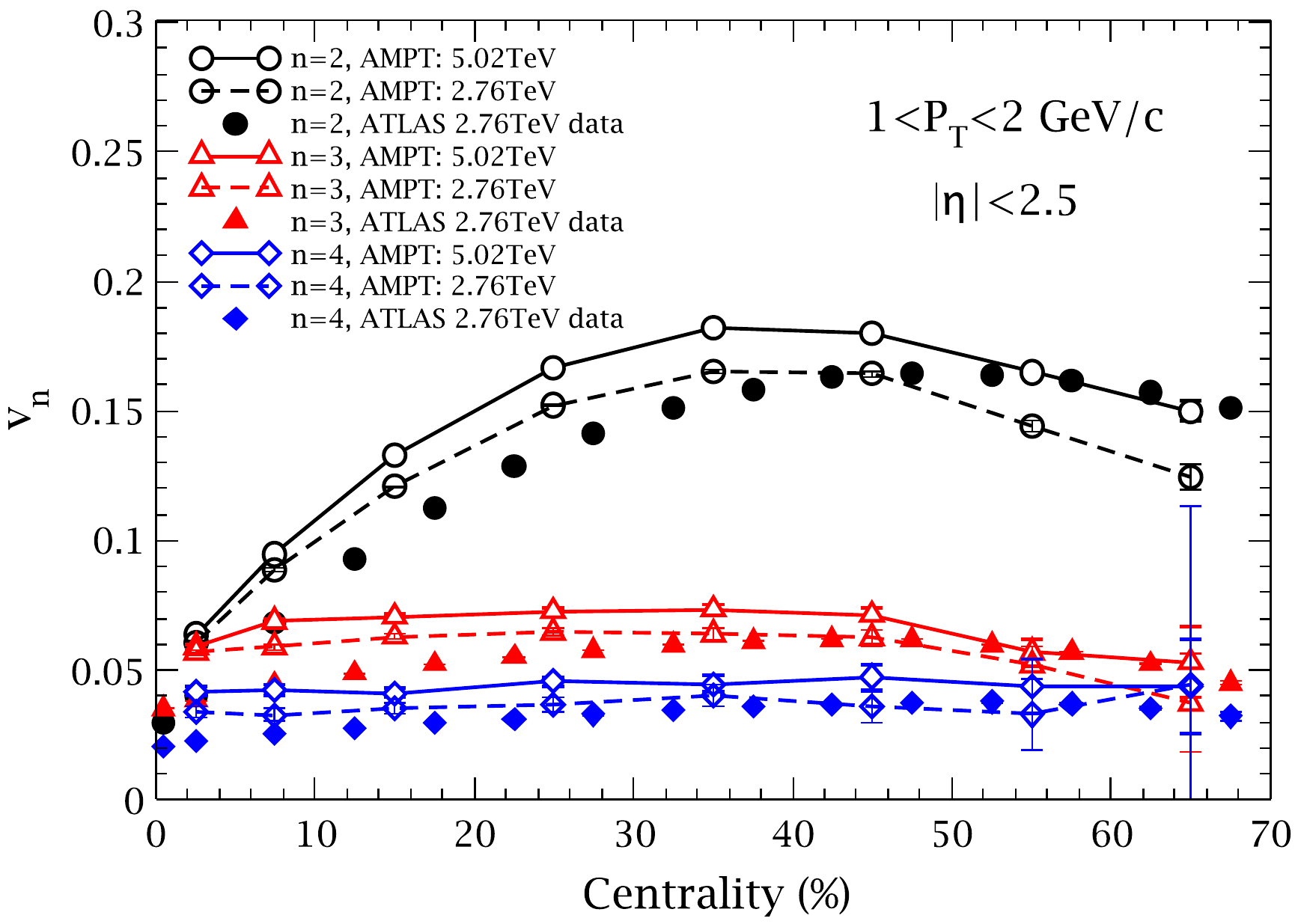}
\caption{Centrality dependences
  of charged particle $v_n (n=2,3,4)$ 
within $1<\pt<2$ GeV/$c$ and $|\eta|<2.5$
from AMPT for Pb+Pb collisions 
in comparison with the ATLAS data at 2.76 TeV.
}
\label{fig:vncent}
\end{figure}

\begin{figure}[h]
\includegraphics[width=6 in]{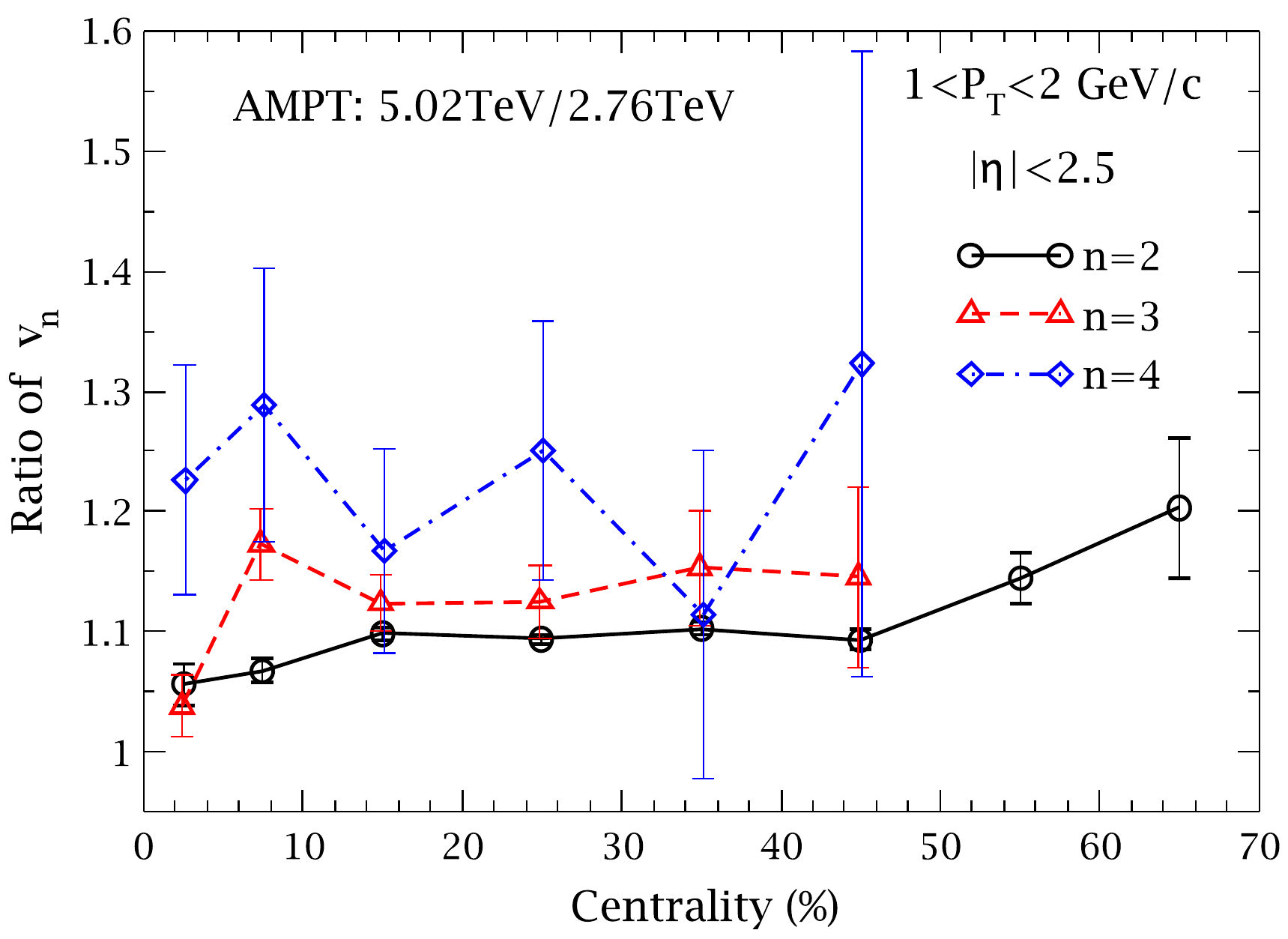}
\caption{
Ratio of the centrality dependence
of charged particle $v_n (n=2,3,4)$ within
$1<\pt<2$ GeV/$c$  and $|\eta|<2.5$ at two different energies from AMPT. 
}
\label{fig:vnratiocent}
\end{figure}

Figure~\ref{fig:vncent} shows the centrality dependences of charged
particle $v_2, v_3$, and $v_4$ within $1<\pt<2$ GeV/$c$ 
and $|\eta|<2.5$ for Pb+Pb collisions at LHC energies. 
From the comparison between the AMPT results at 2.76 TeV (dashed
curves with open symbols) and the corresponding ATLAS data
\cite{ATLAS:2012at}  (filled symbols), 
the AMPT model is seen to reasonably reproduce the overall shapes and 
magnitudes of $v_2, v_3$, and $v_4$. 
We also see that, for each $n$, the AMPT $v_n$ magnitude at 5.02 TeV
is almost always higher than that at 2.76 TeV at the same centrality. 
In Fig.\ref{fig:vnratiocent} we show the ratio of the AMPT $v_n
(n=2,3,4)$ result at 5.02 TeV to that at 2.76 TeV,
where points with very large error bars have been removed 
and the curves for $n=3$ and $n=4$ are slightly shifted horizontally
(by $\pm 0.1\%$) for easier identification. 
We see that the overall relative increase is the smallest for $v_2$
but the largest for $v_4$; this
ordering of the relative $v_n$ increase is the same as that in the
two recent hydrodynamic predictions
\cite{Niemi:2015voa,Noronha-Hostler:2015uye}.
In addition, an overall centrality dependence of the relative increase is 
observed for $v_2$ from the AMPT results, where more peripheral collisions
tend to have a larger relative increase; this is also consistent with the 
two recent hydrodynamic predictions. The magnitudes of the relative
increases in $v_2$ from AMPT as shown in Fig.\ref{fig:vnratiocent} are
bigger than those in the two hydrodynamic predictions
\cite{Niemi:2015voa,Noronha-Hostler:2015uye}, since 
different $\pt$ ranges are used in those calculations.

\begin{figure}[h]
\includegraphics[width=6 in]{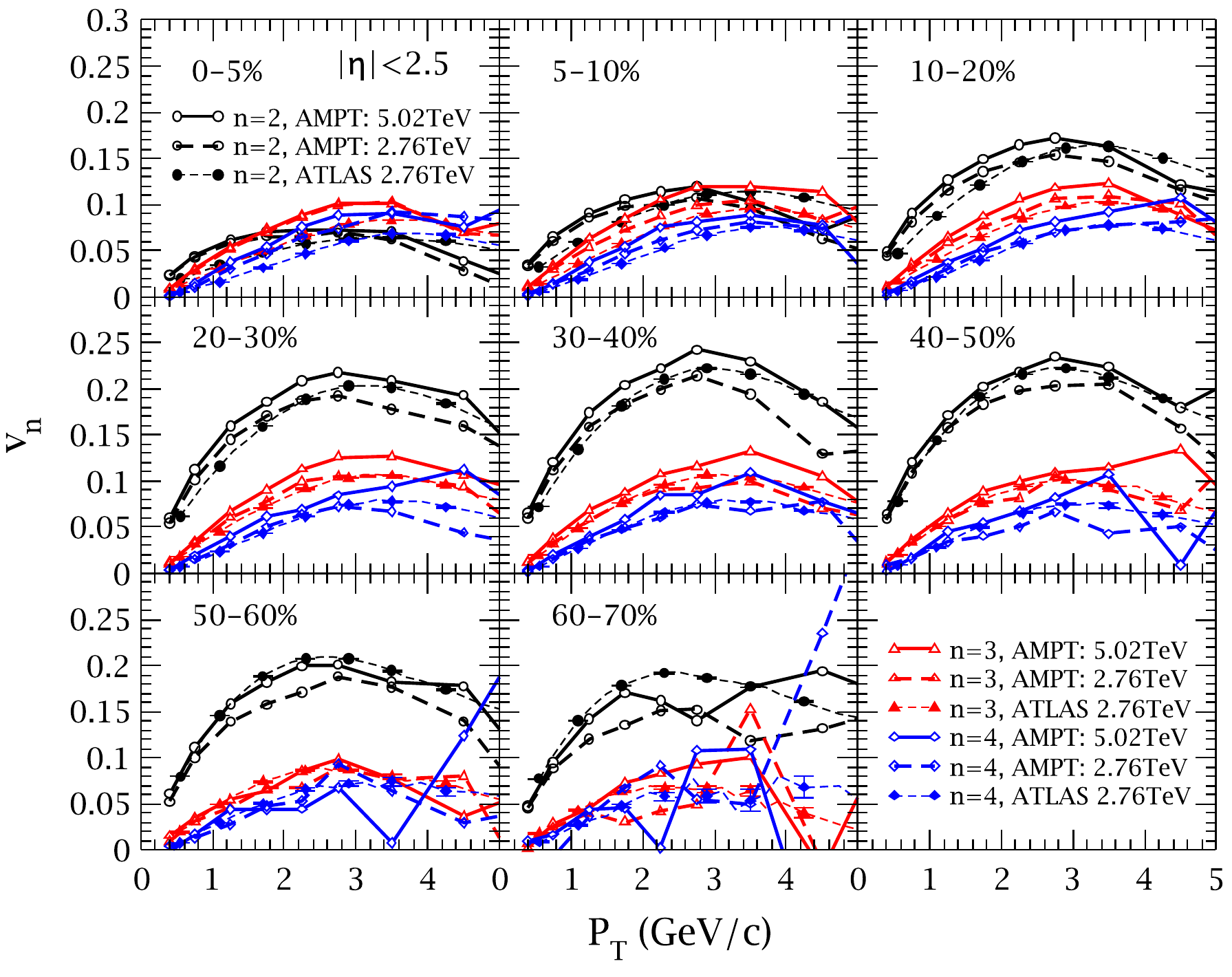}
\caption{The $\pt$ dependences of
  charged particle $v_n (n=2,3,4)$ within $|\eta|<2.5$ at different 
  centralities from AMPT for Pb+Pb collisions in comparison with
  the ATLAS data at 2.76 TeV. 
}
\label{fig:vnptcent}
\end{figure}

AMPT results on the $\pt$ dependences of charged particle $v_2, v_3$,
and $v_4$ within $|\eta|<2.5$ are shown in Fig.\ref{fig:vnptcent}  
for Pb+Pb collisions at 5.02 TeV (solid curves with open symbols) 
and 2.76 TeV (dashed curves with open symbols) over a large
$\pt$ range for eight different centralities. 
The ATLAS data for Pb+Pb collisions at 2.76 TeV \cite{ATLAS:2012at}
(thin dashed curves with filled symbols) are also shown for
comparison.
Overall, the AMPT model reasonably reproduces the shapes and
magnitudes of $v_n$ at 2.76 TeV (especially below $\pt \sim 3$
GeV/$c$), including the ordering of $v_n$ at a given centrality. 
For $v_2$ below $\pt \sim 2$ GeV/$c$, we see that the AMPT model tends
to over-estimate for central collisions but under-estimate for peripheral
collisions, consistent with the centrality dependences of
charged particle $v_2$ shown in Fig.\ref{fig:vncent}. 
On the other hand, AMPT results on the $\pt$ dependences of charged
particle $v_3$ and $v_4$ at low $\pt$ are quite consistent with the data at 2.76
TeV (except for central collisions). 

We have also studied the pseudo-rapidity dependence of
charged particle $v_2$. 
For the $\eta$ dependence, we calculate $v_2$ using the
Event-Plane method similar to the experiments.
We use the $\eta$ range $2.05<|\eta|<3.2$ for 
Au+Au collisions at 200 GeV 
as the PHOBOS experiment and $3<|\eta|<5$ for Pb+Pb collisions at 2.76
TeV as the CMS experiment for the event plane calculations.
Note that the ATLAS range $3.2<|\eta|<4.8$ is similar to the above CMS
range, and for LHC energies we  use the event plane from the $\eta$ side 
opposite to the charged particles used for the $v_2$ calculation. 
Figure~\ref{fig:v2eta} shows the AMPT results for Pb+Pb collisions at
5.02 TeV (solid curves), 2.76 TeV (dashed curves), and Au+Au collisions
at 200 GeV (dot-dashed curves) for two centralities: 
15-25\% and 25-50\%.  
The corresponding experimental data from CMS \cite{Chatrchyan:2012ta} 
(filled circles) and ATLAS \cite{Aad:2014eoa} (open circles) 
for Pb+Pb collisions at 2.76 TeV and from PHOBOS \cite{Back:2004mh}
(open diamonds) for Au+Au collisions at 200 GeV 
are shown for comparison. 
We see that the AMPT model reasonably reproduces the observed
magnitudes and shapes for these collisions at both
centralities.  

Figure~\ref{fig:v2etaratio} shows the ratios of charged particle
$v_2(\eta)$ at one energy to that at a lower energy 
in the overlapping $\eta$ range, including the range $5<|\eta|<6$.
We see that in general the ratio is the lowest at $\eta=0$ and 
gradually increases away from mid-pseudorapidity (at least up to
$|\eta| \sim 2$), and the ratio curves are essentially the same for
the two centrality classes.  
For the ratio of $v_2(\eta)$ in Pb+Pb collisions at 2.76 TeV to that
in Au+Au collisions at 200 GeV, the AMPT results (thick dashed curve 
for the 15-25\% centrality and thin dashed curve for the 25-50\%
centrality) are close to the data, considering the large error bars of
the data.  
For the ratio of $v_2(\eta)$ in Pb+Pb collisions at 5.02 TeV to that at
2.76 TeV (solid curves), the AMPT results show that the overall
magnitudes are lower than the other ratio (dashed curves). 

\subsection{Longitudinal Correlations}

The initial spatial geometry including the event plane depends on
the pseudo-rapidity but also has longitudinal correlations
\cite{Xiao:2012uw,Pang:2014pxa,Pang:2015zrq}. 
This longitudinal correlation comes naturally in the AMPT model
since each wounded nucleon can produce multiple initial particles that
have almost the same transverse position but a range of different $\eta$
values. For the string melting version of AMPT, each wounded nucleon
typically produces many initial partons, therefore the initial
transverse spatial geometry of the parton matter including the event
plane has a strong correlation over a finite $\eta$ range.  
Through partonic and hadronic interactions, the azimuthal
anisotropies $v_n$ will then have correlations over a finite
$\eta$ range.

We follow the definition of the correlation observable
as proposed by the CMS collaboration \cite{Khachatryan:2015oea}.
The pseudo-rapidity range of reference particles is 
$3.0<\eta^b<4.0$ here. 
The factorization ratio $r_{n}(\eta^{a},\eta^{b})$
is calculated as 
\begin{equation}
\label{rn_eta}
r_{n}(\eta^{a},\eta^{b}) \equiv
\frac{V_{n\Delta}(-\eta^{a},\eta^{b})}{V_{n\Delta}(\eta^{a},\eta^{b})}, 
\end{equation}
where $V_{n\Delta}(\eta^{a},\eta^{b})$ is defined as
$\langle\langle\cos(n\Delta\phi) \rangle\rangle_{S}-\langle\langle
\cos (n\Delta\phi)\rangle\rangle_{B}$. Here, $\langle\langle
\rangle\rangle$ denotes the averaging over all particle pairs in each
event and over all the events. The subscript $S$ corresponds to the
average over pairs taken from the same event, while $B$ represents the
mixing of particles from two randomly-selected events from the same
centrality class.

\begin{figure}[h]
\includegraphics[width=6 in]{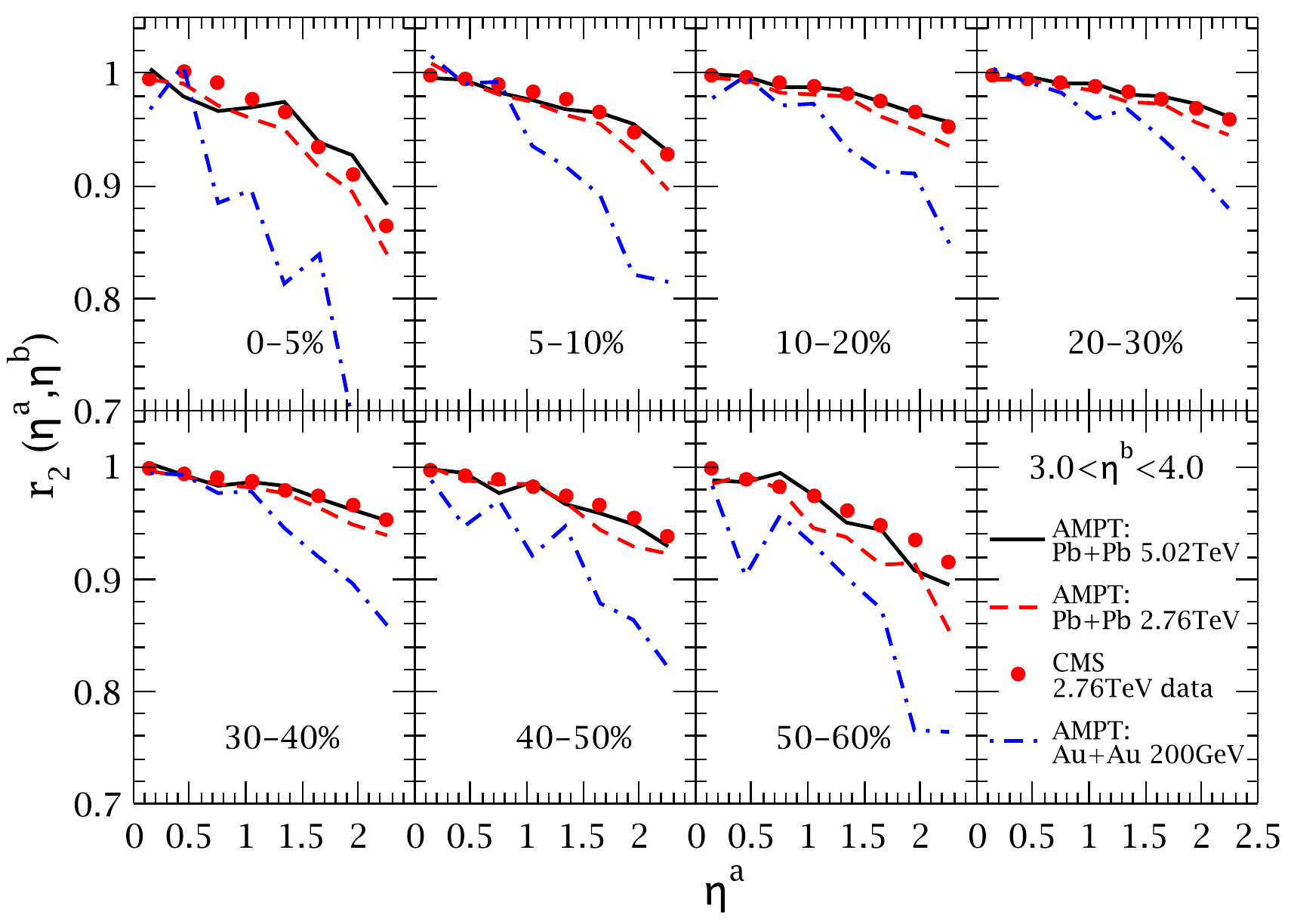}
\caption{AMPT results on the factorization ratio 
  $r_{2}(\eta^{a},\eta^{b})$ as functions of $\eta^a$ in comparison
  with the CMS 2.76 TeV data for different centralities. 
}
\label{fig:r2}
\end{figure}

AMPT results on the factorization ratio $r_{2}(\eta^{a},\eta^{b})$ as
a function of $\eta^a$ are shown in Fig.\ref{fig:r2} for Pb+Pb
collisions at  5.02 TeV (solid curves), 2.76 TeV (dashed curves), and
Au+Au collisions at 200 GeV (dot-dashed curves) for seven different
centralities.  We see that the AMPT results at 2.76 TeV are rather
consistent with  the corresponding CMS data
\cite{Khachatryan:2015oea},  similar to an earlier study
\cite{Pang:2015zrq} that used the string melting AMPT model as the
initial condition for an ideal (3+1)D hydrodynamics. 
Furthermore, the AMPT results show that the longitudinal correlation
is much suppressed in pseudo-rapidity for Au+Au collisions at 200 GeV,
again similar to the earlier study \cite{Pang:2015zrq}. 

\begin{figure}[h]
\includegraphics[width=6 in]{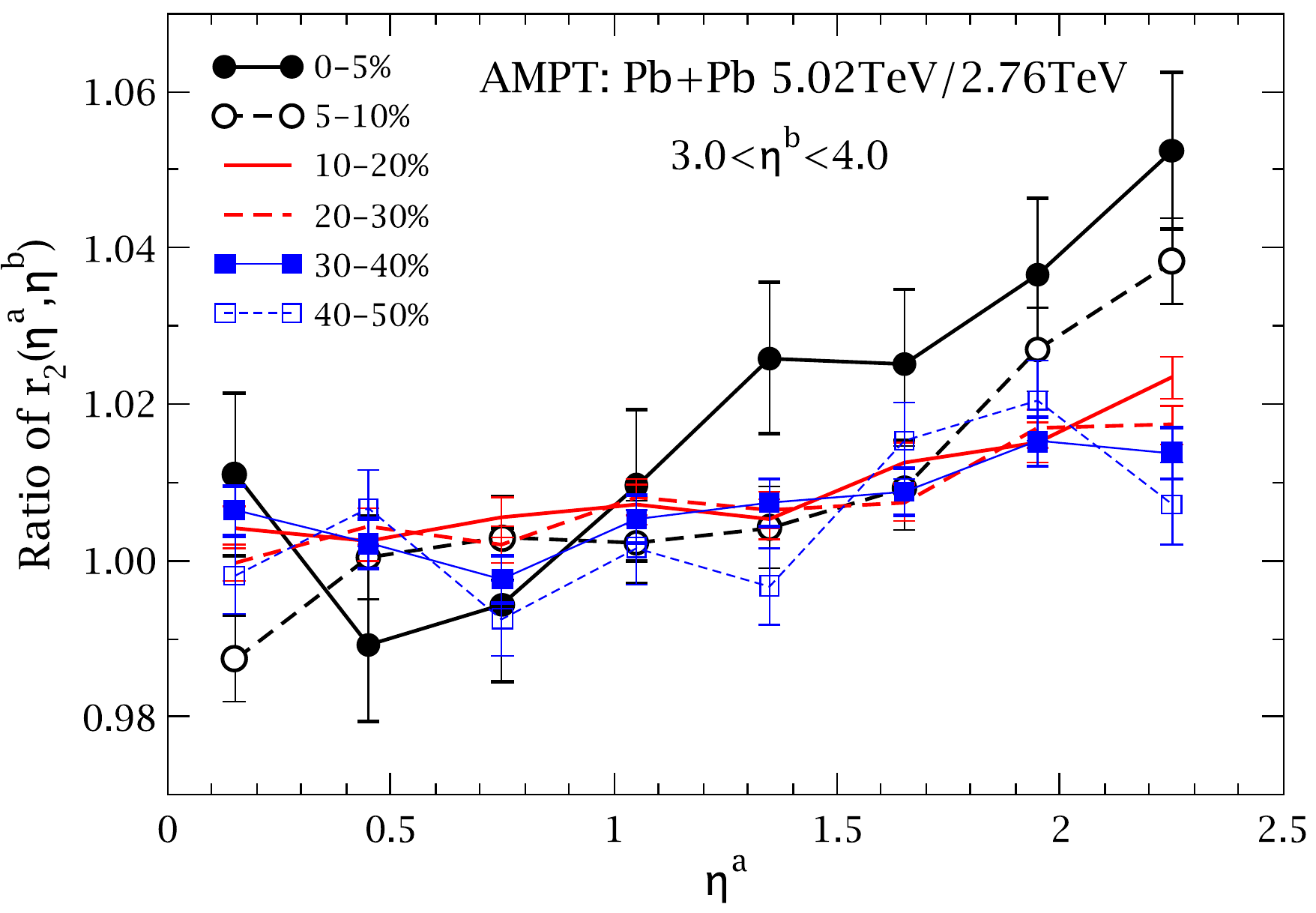}
\caption{Ratio of the factorization ratio
  $r_{2}(\eta^{a},\eta^{b})$ at 5.02 TeV to that at 2.76 TeV  
as a function of $\eta^a$ for different centralities of Pb+Pb collisions.
}
\label{fig:r2ratio}
\end{figure}

Comparing AMPT results at 5.02 TeV with those at 2.76 TeV, we see that
the longitudinal correlation at 5.02 TeV is slightly stronger. 
This can be better seen in Fig.\ref{fig:r2ratio}, which shows the ratios of
$r_{2}(\eta^{a},\eta^{b})$ at 5.02 TeV to   
that at 2.76TeV for most centrality classes. 
In addition, Fig.\ref{fig:r2ratio} shows that 
the increase of the ratio with $\eta^a$ systematically depends on the
centrality, with central collisions having the strongest relative 
increase of $r_{2}(\eta^{a},\eta^{b})$ with $\eta^a$. 

\begin{figure}[h]
\includegraphics[width=6 in]{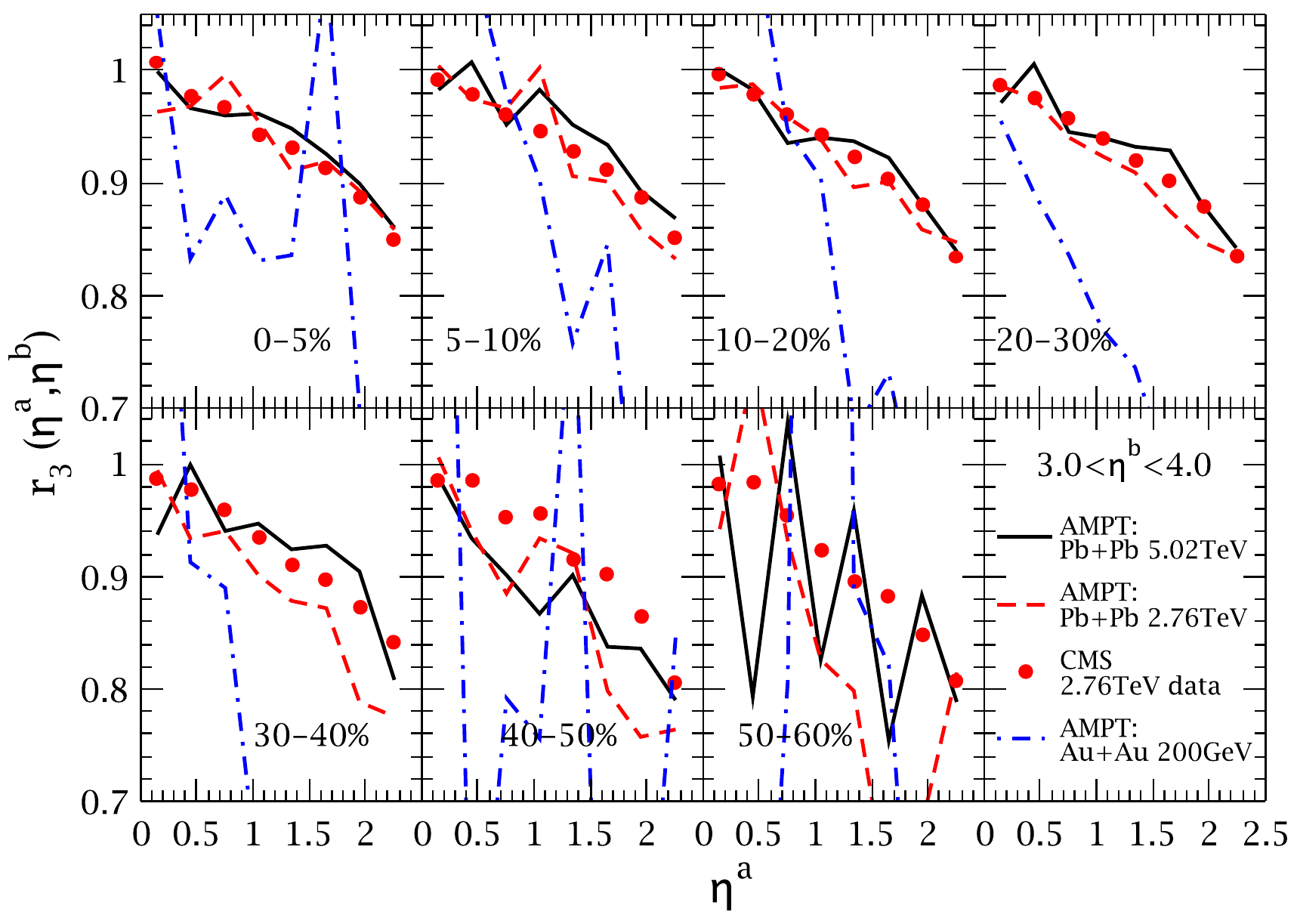}
\caption{
Same as Fig.\ref{fig:r2} but for factorization ratio $r_{3}(\eta^{a},\eta^{b})$.
}
\label{fig:r3}
\end{figure}

AMPT results on the factorization ratio $r_{3}(\eta^{a},\eta^{b})$ 
are shown in Fig.\ref{fig:r3} 
in comparison with the CMS data for Pb+Pb collisions at 2.76 TeV
\cite{Khachatryan:2015oea}.
We see that the AMPT results at 2.76 TeV are roughly consistent with 
data and that the longitudinal correlation
is much suppressed for Au+Au collisions at 200 GeV,  
qualitatively similar to an earlier study \cite{Pang:2015zrq}. 
In addition, the factorization ratio $r_{3}(\eta^{a},\eta^{b})$ 
at 5.02 TeV is mostly somewhat higher than 
that at 2.76 TeV at the same centrality.
However, statistical error bars of the AMPT
$r_{3}(\eta^{a},\eta^{b})$  results are often too big for us to draw
more conclusions. 

\section{Discussions}

We have seen that the string melting version of the AMPT model 
can describe the qualitative features of many observables, 
and it can often reasonably describe the experimental data
quantitatively. 
However, we have also seen that it is sometimes inconsistent with
data, for example, the centrality dependence of variables such as  
the mid-pseudorapidity charged particle yields 
as shown in Figs.\ref{fig:dnchde} and \ref{fig:dnchderatio} and $v_n (n=2,3,4)$
as shown in Figs.\ref{fig:vncent} and \ref{fig:vnptcent}. 

\begin{figure}[h]
\includegraphics[width=6 in]{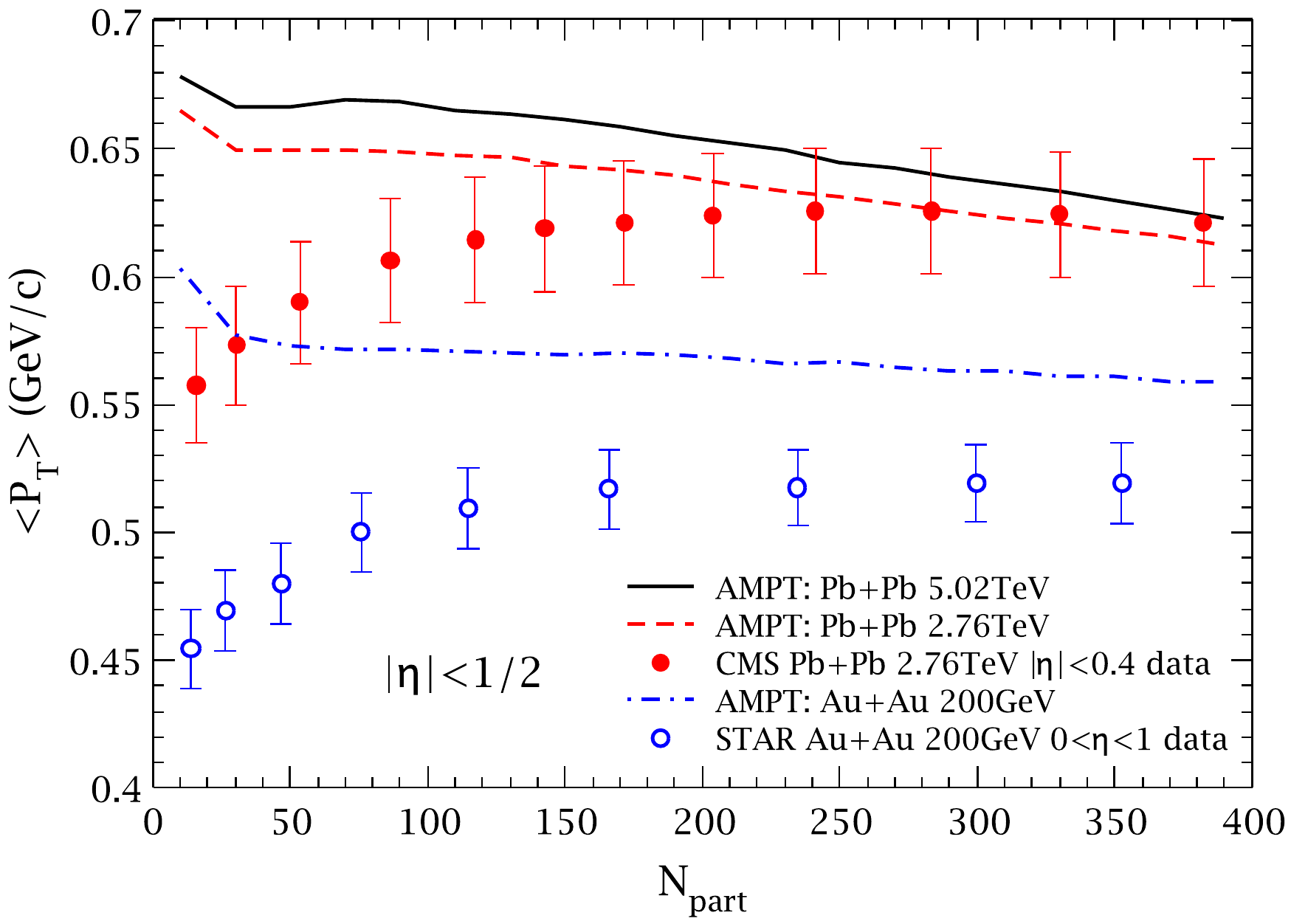}
\caption{AMPT results on the centrality dependence
  of charged particle mean transverse momentum, 
 $\langle \pt \rangle$, at mid-pseudorapidity in comparison with the
 CMS data at 2.76 TeV and STAR data at 200 GeV.
}
\label{fig:meanpt}
\end{figure}

Figure~\ref{fig:meanpt} shows that the centrality dependence of
charged particle mean transverse momentum $\langle \pt \rangle$ at
mid-pseudorapidity from the AMPT is also inconsistent with the CMS
data at 2.76 TeV \cite{Chatrchyan:2012ta} and STAR data at 200 GeV
\cite{Adams:2004cb}.  
Experimental data show an overall increase of $\langle \pt \rangle$ 
with $\np$, while the string melting AMPT results show an overall
decrease. It was known that the string melting version of AMPT
leads to a smaller $\langle \pt \rangle$ for more central
collisions, while the default version of AMPT shows the
opposite \cite{Haque:2011aa}. 
One reason is that in the string melting AMPT model each parton scatters
more frequently on the average for more central collisions, 
and more collisions lead to a bigger decrease of the parton $\langle
\pt \rangle$. 
Similar to earlier AMPT studies, in this study we use the same Lund
string fragmentation $a$ and $b$ values for different centralities at
a given energy. 
On the other hand, there is considerable uncertainty regarding the
initial condition of heavy ion collisions, and the Lund $a$ and $b$
values, which controls the initial parton production including the
initial  parton $\langle \pt \rangle$, may
depend on centrality and/or system size \cite{Lin:2004en}.  
This will consequently affect the centrality dependence of observables
calculated from the AMPT model.

\begin{figure}[h]
\includegraphics[width=6 in]{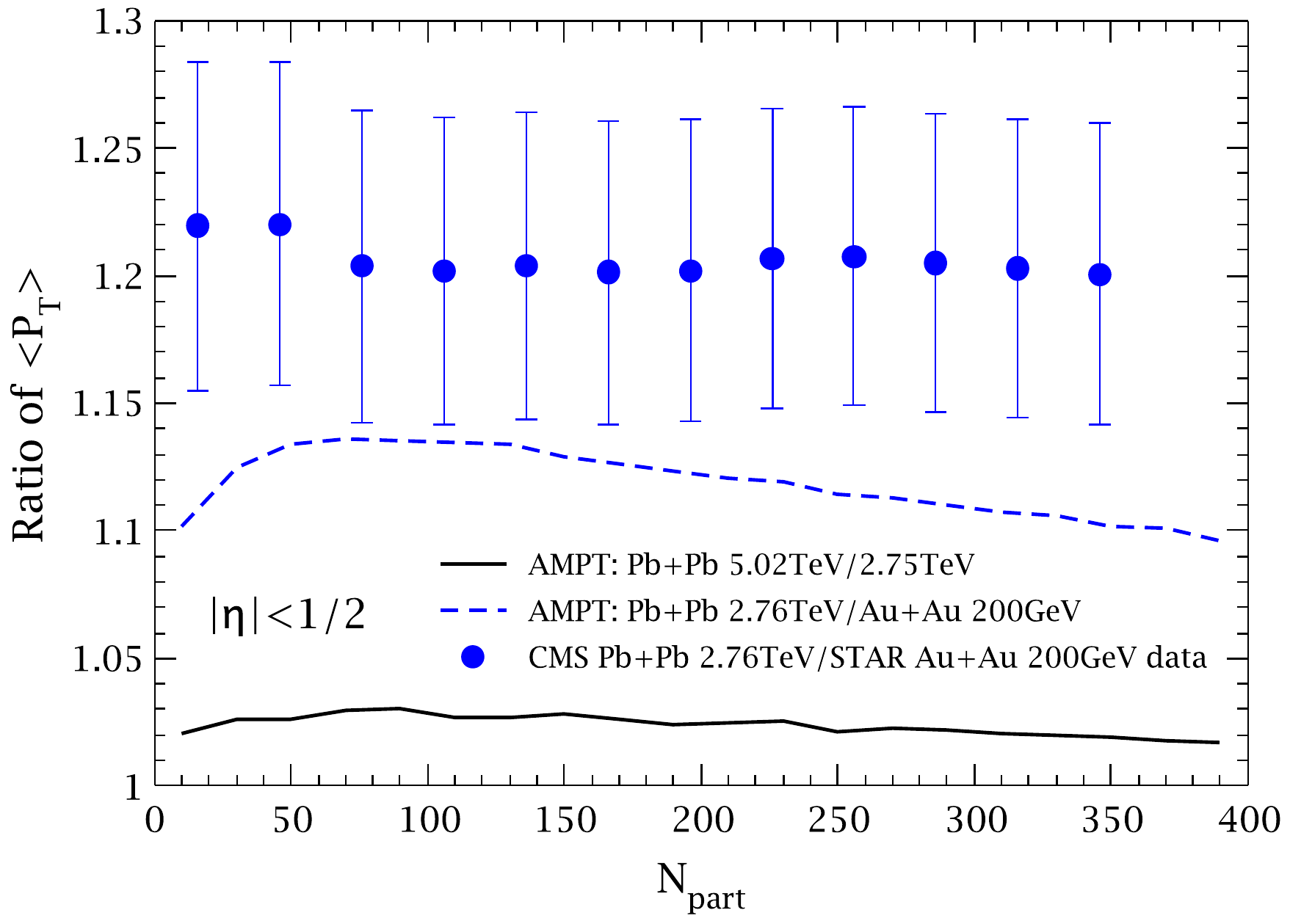}
\caption{Ratio of 
 charged particle $\langle \pt \rangle$ at mid-pseudorapidity 
at two different energies as a function of $\np$ from AMPT in
comparison with data.  
}
\label{fig:meanptratio}
\end{figure}

Since we may expect some uncertainties to cancel out in ratios, 
we show in Fig.\ref{fig:meanptratio} the ratio of 
the charged particle $\langle \pt \rangle$ at mid-pseudorapidity 
at one energy to that at a lower energy. 
The ratio of the AMPT $\langle \pt \rangle$ at 2.76 TeV to that at 200
GeV (dashed curve) is seen to have a weak centrality
dependence similar to the data; while the overall
magnitude of the AMPT ratio is lower than the data. 
We also see from the AMPT results that, as the energy of Pb+Pb
collisions increases from 2.76 TeV to 5.02 TeV, $\langle \pt
\rangle$ at mid-pseudorapidity only increases slightly (between 1.7\%
and 3.0\%). The overall magnitude of the $\langle \pt
\rangle$ relative increase from AMPT is close to that 
from a recent hydrodynamic prediction \cite{Noronha-Hostler:2015uye}.
However the centrality dependences are different: 
AMPT results show a small decrease of the $\langle \pt \rangle$  ratio going
from mid-central towards central collisions, while the hydrodynamic
prediction \cite{Noronha-Hostler:2015uye} shows a small increase.

\begin{figure}[h]
\includegraphics[width=6 in]{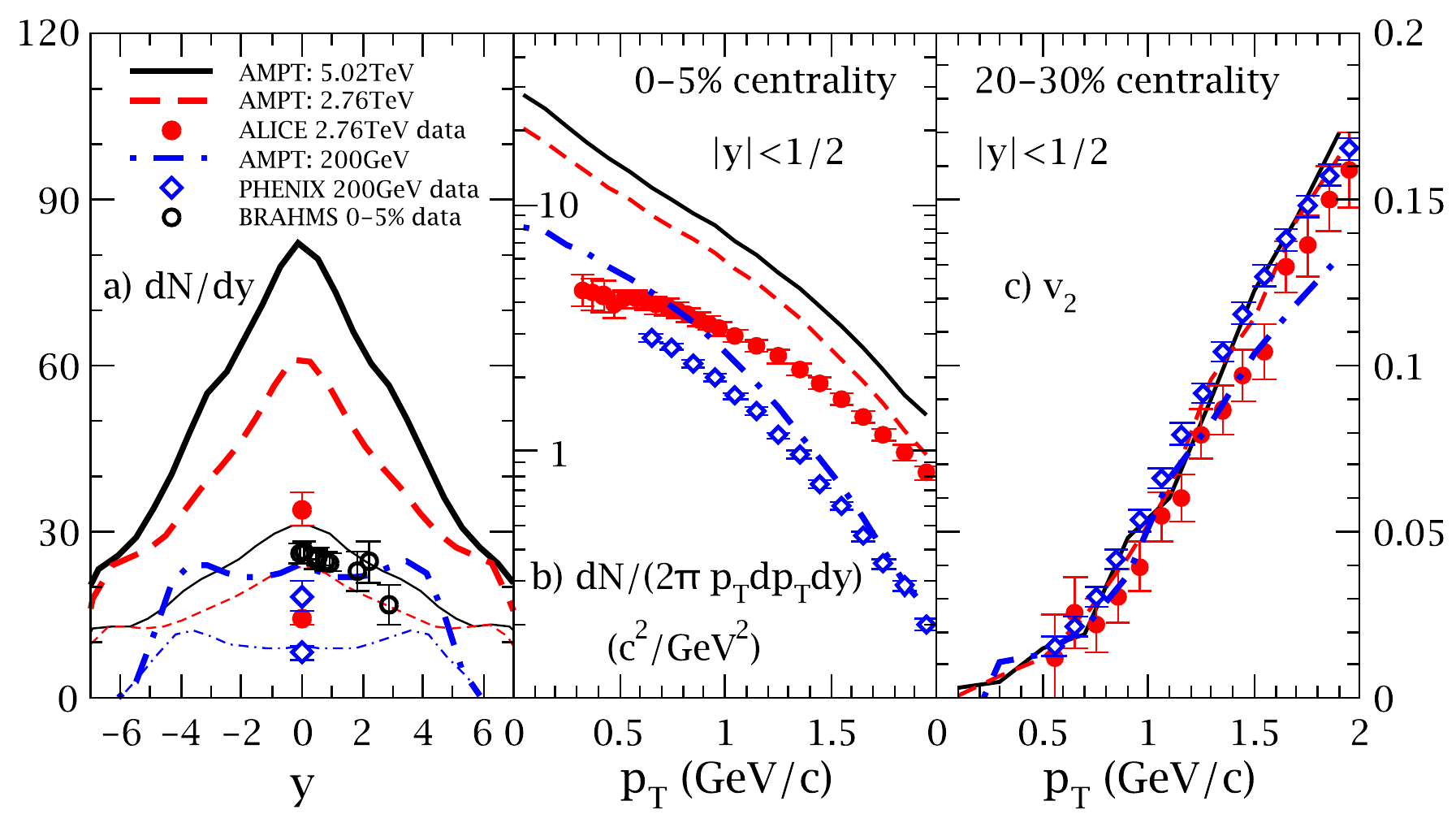}
\caption{AMPT results on protons 
in Pb+Pb collisions at 5.02 TeV, 2.76 TeV, and Au+Au collisions at 200 GeV:
a) dN/dy for 0-5\% and 20-30\% centralities, 
b) $\pt$ spectra at mid-rapidity for the 0-5\% centrality, 
and c) $v_2$ as a function of $\pt$  at mid-rapidity for the 20-30\%
centrality. Corresponding experimental data at 2.76 TeV and 200 GeV
are also shown for comparison. 
}
\label{fig:proton}
\end{figure}

Proton rapidity distribution is another observable that the string melting
version of the AMPT model fails to describe. It was realized 
earlier on \cite{Lin:2004en} that the string melting version of AMPT
(unlike the default version of AMPT)  
gives an artificial peak at mid-rapidity in the proton and
anti-proton rapidity distributions for central Au+Au collisions at 200
GeV, which points to the need to improve the simple quark coalescence
model in AMPT \cite{Lin:2011zzg}.
As shown in Fig.\ref{fig:proton}a, the string melting AMPT model 
also gives mid-rapidity peaks in both central (thick curves) and
semi-central (thin curves) Pb+Pb collisions at LHC energies.
In addition, the model significantly over-estimates the proton dN/dy
at mid-rapidity in comparison with the corresponding ALICE data at
2.76 TeV (filled circles).
Figure~\ref{fig:proton}b shows the proton $\pt$ spectra at
mid-rapidity for central collisions, where we see that 
the AMPT model mostly over-estimates at low $\pt$ and the
over-estimation factor is bigger at 2.76 TeV than that at 200 GeV.
On the other hand, the elliptic flow is normalized by the particle
multiplicity, therefore it is not directly affected by this over-estimation. 
Figure~\ref{fig:proton}c indeed shows that the proton $v_2$ results at
mid-rapidity from AMPT (curves without
symbols) agree reasonably well with the ALICE data at 2.76 TeV and
PHENIX data at 200 GeV  for the 20-30\% centrality,  
where the AMPT $v_2$ results at 200 GeV has the cut $|\eta|<0.35$
for comparison with the PHENIX data. 
Nevertheless, it is important to address the problem in baryon
rapidity distributions in the string melting version of AMPT. 
Improvements of the quark coalescence model in AMPT will be needed;
for example, it has been proposed that one should use the local energy 
density as the criterion of hadronization in order to make the
effective equation of state more realistic
\cite{Lin:2004en,Lin:2011zzg,Zhang:2015tta}.

\section{Summary}

Using the string melting version of a multi-phase transport model with 
parameters established from an earlier study, 
we present predictions on various observables in Pb+Pb collisions at
$\snn=5.02$ TeV. 
AMPT results and experimental data for Pb+Pb collisions at 2.76 TeV
and Au+Au collisions at 200 GeV are also often shown for comparisons. 
We first compare with the already-available centrality dependence data
on charged particle $dN/d\eta$ in Pb+Pb collisions at 5.02 TeV. 
We then present AMPT model predictions on identified particle dN/dy,
$\pt$-spectra at mid-rapidity for central collisions, 
azimuthal anisotropies $v_n (n=2,3,4)$ including the $\pt$, centrality
and $\eta$ dependences, and factorization ratios
$r_{n}(\eta^{a},\eta^{b}) (n=2,3)$ for longitudinal correlations.  

\section{Acknowledgments}
G.-L. M. is supported by the Major State Basic Research Development Program in China under Grant No. 2014CB845404, the National Natural Science Foundation of China under Grants No. 11522547, 11375251, and 11421505, the Youth Innovation Promotion Association of CAS under Grant No. 2013175.


\begin{thebibliography}{99}

\bibitem{Huovinen:2001cy} 
  P.~Huovinen, P.~F.~Kolb, U.~W.~Heinz, P.~V.~Ruuskanen and S.~A.~Voloshin,
  Phys.\ Lett.\ B {\bf 503}, 58 (2001).

\bibitem{Betz:2008ka} 
  B.~Betz, J.~Noronha, G.~Torrieri, M.~Gyulassy, I.~Mishustin and D.~H.~Rischke,
  Phys.\ Rev.\ C {\bf 79}, 034902 (2009).

\bibitem{Schenke:2010rr} 
  B.~Schenke, S.~Jeon and C.~Gale,
  Phys.\ Rev.\ Lett.\  {\bf 106}, 042301 (2011).

\bibitem{Xu:2004mz} 
  Z.~Xu and C.~Greiner,
  Phys.\ Rev.\ C {\bf 71}, 064901 (2005).

\bibitem{Lin:2004en} 
  Z.~W.~Lin, C.~M.~Ko, B.~A.~Li, B.~Zhang and S.~Pal,
  Phys.\ Rev.\ C {\bf 72}, 064901 (2005).

\bibitem{Cassing:2009vt} 
  W.~Cassing and E.~L.~Bratkovskaya,
  Nucl.\ Phys.\ A {\bf 831}, 215 (2009).

\bibitem{Petersen:2008dd} 
  H.~Petersen, J.~Steinheimer, G.~Burau, M.~Bleicher and H.~Stocker,
  Phys.\ Rev.\ C {\bf 78}, 044901 (2008).

\bibitem{Werner:2010aa} 
  K.~Werner, I.~Karpenko, T.~Pierog, M.~Bleicher and K.~Mikhailov,
  Phys.\ Rev.\ C {\bf 82}, 044904 (2010).

\bibitem{Song:2010mg} 
  H.~Song, S.~A.~Bass, U.~Heinz, T.~Hirano and C.~Shen,
  Phys.\ Rev.\ Lett.\  {\bf 106}, 192301 (2011); 
{\bf 109}, 139904 (2012).

\bibitem{Bozek:2011if} 
  P.~Bozek,
  Phys.\ Rev.\ C {\bf 85}, 014911 (2012).

\bibitem{Ma:2014pva} 
  G.~L.~Ma and A.~Bzdak,
  Phys.\ Lett.\ B {\bf 739}, 209 (2014).

\bibitem{Bzdak:2014dia} 
  A.~Bzdak and G.~L.~Ma,
  Phys.\ Rev.\ Lett.\  {\bf 113}, 252301 (2014).

\bibitem{He:2015hfa} 
  L.~He, T.~Edmonds, Z.~W.~Lin, F.~Liu, D.~Molnar and F.~Wang,
  Phys.\ Lett.\ B {\bf 753}, 506 (2016).

\bibitem{Lin:2015ucn} 
  Z.~W.~Lin, L.~He, T.~Edmonds, F.~Liu, D.~Molnar and F.~Wang,
  arXiv:1512.06465 [nucl-th].

\bibitem{Niemi:2015voa} 
  H.~Niemi, K.~J.~Eskola, R.~Paatelainen and K.~Tuominen,
  Phys.\ Rev.\ C {\bf 93}, 014912 (2016).

\bibitem{Noronha-Hostler:2015uye} 
  J.~Noronha-Hostler, M.~Luzum and J.~Y.~Ollitrault,
  Phys.\ Rev.\ C {\bf 93}, 034912 (2016).

\bibitem{Zhang:1999bd} 
  B.~Zhang, C.~M.~Ko, B.~A.~Li and Z.~W.~Lin,
  Phys.\ Rev.\ C {\bf 61}, 067901 (2000).

\bibitem{Bass:1999zq} 
  S.~A.~Bass {\it et al.},
  Nucl.\ Phys.\ A {\bf 661}, 205 (1999).

\bibitem{Lin:2000cx} 
  Z.~W.~Lin, S.~Pal, C.~M.~Ko, B.~A.~Li and B.~Zhang,
  Phys.\ Rev.\ C {\bf 64}, 011902(R) (2001).

\bibitem{Gyulassy:1994ew} 
  M.~Gyulassy and X.~N.~Wang,
  Comput.\ Phys.\ Commun.\  {\bf 83}, 307 (1994).

\bibitem{Zhang:1997ej} 
  B.~Zhang,
  Comput.\ Phys.\ Commun.\  {\bf 109}, 193 (1998).

\bibitem{Sjostrand:1993yb} 
  T.~Sjostrand,
  Comput.\ Phys.\ Commun.\  {\bf 82}, 74 (1994).

\bibitem{Li:1995pra} 
  B.~A.~Li and C.~M.~Ko,
  Phys.\ Rev.\ C {\bf 52}, 2037 (1995).

\bibitem{Lin:2001zk} 
  Z.~W.~Lin and C.~M.~Ko,
  Phys.\ Rev.\ C {\bf 65}, 034904 (2002).

\bibitem{Lin:2002gc} 
  Z.~W.~Lin, C.~M.~Ko and S.~Pal,
  Phys.\ Rev.\ Lett.\  {\bf 89}, 152301 (2002).

\bibitem{Abreu:2007kv} 
  N.~Armesto {\it et al.},
  J.\ Phys.\ G {\bf 35}, 054001 (2008).

\bibitem{Lin:2014tya} 
  Z.~W.~Lin,
  Phys.\ Rev.\ C {\bf 90}, 014904 (2014).

\bibitem{ampt} 
AMPT source files are available at http://myweb.ecu.edu/linz/ampt/

\bibitem{Adam:2015ptt} 
  J.~Adam {\it et al.} [ALICE Collaboration],
  arXiv:1512.06104 [nucl-ex].

\bibitem{Aamodt:2010cz} 
  K.~Aamodt {\it et al.} [ALICE Collaboration],
  Phys.\ Rev.\ Lett.\  {\bf 106}, 032301 (2011).

\bibitem{Adler:2004zn} 
  S.~S.~Adler {\it et al.} [PHENIX Collaboration],
  Phys.\ Rev.\ C {\bf 71}, 034908 (2005).

\bibitem{Abbas:2013bpa} 
  E.~Abbas {\it et al.} [ALICE Collaboration],
  Phys.\ Lett.\ B {\bf 726}, 610 (2013).

\bibitem{Adam:2015kda} 
  J.~Adam {\it et al.} [ALICE Collaboration],
  Phys.\ Lett.\ B {\bf 754}, 373 (2016).

\bibitem{Abelev:2013vea} 
  B.~Abelev {\it et al.} [ALICE Collaboration],
  Phys.\ Rev.\ C {\bf 88}, 044910 (2013).

\bibitem{Adler:2003cb} 
  S.~S.~Adler {\it et al.} [PHENIX Collaboration],
  Phys.\ Rev.\ C {\bf 69}, 034909 (2004).

\bibitem{Bearden:2004yx} 
  I.~G.~Bearden {\it et al.} [BRAHMS Collaboration],
  Phys.\ Rev.\ Lett.\  {\bf 94}, 162301 (2005).

\bibitem{Poskanzer:1998yz} 
  A.~M.~Poskanzer and S.~A.~Voloshin,
  Phys.\ Rev.\ C {\bf 58}, 1671 (1998).

\bibitem{Chatrchyan:2013nka} 
  S.~Chatrchyan {\it et al.} [CMS Collaboration],
  Phys.\ Lett.\ B {\bf 724}, 213 (2013).

\bibitem{ATLAS:2012at} 
G.~Aad {\it et al.} [ATLAS Collaboration],
Phys.\ Rev.\ C {\bf 86}, 014907 (2012).

\bibitem{Gu:2012br} 
Y.~Gu [PHENIX Collaboration],
Nucl.\ Phys.\ A {\bf 904-905}, 353c (2013).

\bibitem{Chatrchyan:2012ta}    
S.~Chatrchyan {\it et al.} [CMS Collaboration],   
  Phys.\ Rev.\ C {\bf 87}, 014902 (2013).

\bibitem{Aad:2014eoa} 
  G.~Aad {\it et al.} [ATLAS Collaboration],
  Eur.\ Phys.\ J.\ C {\bf 74}, 2982 (2014).

\bibitem{Back:2004mh}    
B.~B.~Back {\it et al.} [PHOBOS Collaboration],   
  Phys.\ Rev.\ C {\bf 72}, 051901(R) (2005).

\bibitem{Xiao:2012uw} 
  K.~Xiao, F.~Liu and F.~Wang,
  Phys.\ Rev.\ C {\bf 87}, 011901(R) (2013).

\bibitem{Pang:2014pxa} 
  L.~G.~Pang, G.~Y.~Qin, V.~Roy, X.~N.~Wang and G.~L.~Ma,
  Phys.\ Rev.\ C {\bf 91}, 044904 (2015).

\bibitem{Pang:2015zrq} 
  L.~G.~Pang, H.~Petersen, G.~Y.~Qin, V.~Roy and X.~N.~Wang,
  Eur.\ Phys.\ J.\ A {\bf 52}, 97 (2016).

\bibitem{Khachatryan:2015oea} 
  V.~Khachatryan {\it et al.} [CMS Collaboration],
  Phys.\ Rev.\ C {\bf 92}, 034911 (2015).
  
\bibitem{Adams:2004cb} 
  J.~Adams {\it et al.} [STAR Collaboration],
  Phys.\ Rev.\ C {\bf 70}, 054907 (2004).

\bibitem{Haque:2011aa} 
  M.~R.~Haque, Z.~W.~Lin and B.~Mohanty,
  Phys.\ Rev.\ C {\bf 85}, 034905 (2012).

\bibitem{Lin:2011zzg} 
  Z.~W.~Lin,
  Indian J.\ Phys.\  {\bf 85}, 837 (2011).

\bibitem{Zhang:2015tta} 
  Y.~Zhang, J.~Zhang, J.~Liu and L.~Huo,
  Phys.\ Rev.\ C {\bf 92}, 014909 (2015).

\end{thebibliography}
\end{document}